\documentclass[11pt]{article}
\usepackage{jcappub}
\usepackage{xcolor}
\usepackage{amsfonts, amsmath}
\usepackage{booktabs}
\usepackage[english]{babel}
\usepackage{colortbl}
\usepackage{epsfig}
\usepackage{epstopdf}
\usepackage{graphicx}
\usepackage{subcaption}
\usepackage{lipsum}
\usepackage{hyperref}
\usepackage{bm}
\usepackage{tikz}
\usetikzlibrary{decorations.markings}
\usepackage{physics}
\usepackage{slashed}
\usepackage[load-configurations=astronomy, range-units=brackets, range-phrase=-, per-mode=reciprocal, mode=math]{siunitx}

\definecolor{lightblue}{rgb}{0.368417,0.506779,0.709798}
\definecolor{lightorange}{rgb}{0.880722,0.611041,0.142051}
\definecolor{lightred}{rgb}{0.922526,0.385626,0.209179}
\definecolor{lightgray}{rgb}{0.5,0.5,0.5}
\definecolor{lightgreen}{rgb}{0.560181,0.691569,0.194885}

\def\kMpc{\, h \, {\rm Mpc}^{-1}}

\begin{document}

\title{Dark Matter Recoupling}

\author[a,b,*]{Eugenia Dallari,}
\author[c,d,*]{Francesco Castagna,}
\author[c,d]{Emanuele Castorina,}
\author[c,d]{Maria Archidiacono,}
\author[a,b]{Ennio Salvioni$\,$}
\affiliation[a]{$\,$Departament de F\'isica, Universitat Aut\`onoma de Barcelona, 08193 Bellaterra, Barcelona, Spain}
\affiliation[b]{$\,$IFAE and BIST, Campus UAB, 08193 Bellaterra, Barcelona, Spain}
\affiliation[c]{$\,$Dipartimento di Fisica ``Aldo Pontremoli'', Universit\`a degli Studi di Milano,\\ Via Celoria 16, 20133 Milan, Italy}
\affiliation[d]{$\,$INFN, Sezione di Milano, Via Celoria 16, 20133 Milan, Italy}
\affiliation[*]{{\it Contributed equally}}

\date{\today}

%\preprint{}

\emailAdd{edallari@ifae.es, castagnaf98@gmail.com, emanuele.castorina@unimi.it, maria.archidiacono@unimi.it,
esalvioni@ifae.es}

\abstract{In the late Universe, and on cosmological scales, dark matter is conventionally assumed to be collisionless, as a consequence of the strong existing bounds on dark matter interactions at the Cosmic Microwave Background last-scattering surface. Challenging this lore, here we show that dark matter interactions can be naturally weak at early times, but then grow to observationally relevant strengths at very late times, even significantly after reionization. This is realized if dark matter {\it recouples} to a dark radiation species in the range of redshifts probed by the current generation of galaxy surveys. We systematically study, for the first time, the phenomenology of this dark matter recoupling scenario. A combination of Cosmic Microwave Background and Baryon Acoustic Oscillation data show that the interaction needs to be weak at present, if the entirety of dark matter couples to dark radiation. From a complementary perspective, we estimate that a $\sim 4\%$ fraction of dark matter could still be strongly interacting today. Implications for a microscopic model realizing the recoupling dynamics are discussed.}

\maketitle

\section{Introduction}
\label{sec:intro}
The collisionless nature of Dark Matter (DM) is a main assumption of the standard cosmological model. Scattering of DM with other species would lead to transfer of momentum and thus to a suppression of the growth of DM fluctuations, which in turn can affect astrophysical and cosmological observations. Depending on the size and redshift dependence of the associated momentum transfer rate, different possible interactions between DM and other particles would become relevant at different stages of the evolution of the Universe, and can thus be probed by different observables.

Significant scattering of DM with Standard Model (SM) particles is viable only at very high redshift, typically before hydrogen recombination. In such models, DM interactions suppress the growth of matter perturbations for the modes that enter the horizon deep in radiation domination, leading to constraints from the anisotropies in the Cosmic Microwave Background (CMB), or, at late times, from the transmitted flux of the Lyman-$\alpha$ forest spectrum of distant quasars and from the distribution of satellite galaxies in the Milky Way~\cite{Boehm:2000gq,Chen:2002yh,Sigurdson:2004zp,Boehm:2004th,Dvorkin:2013cea,Ali-Haimoud:2015pwa,Gluscevic:2017ywp,Slatyer:2018aqg,Boddy:2018wzy,Nadler:2019zrb,Maamari:2020aqz,Becker:2020hzj,Buen-Abad:2021mvc,Nguyen:2021cnb,Rogers:2021byl,Boddy:2022tyt,He:2025npy}. Moreover,  
such interactions mostly lead to decoupling over time, i.e.,~as the Universe expands scattering events become less frequent, making them completely irrelevant in the low-redshift Universe. Scenarios where DM interacts with baryons with a cross section proportional to $v^{-4}$, where $v$ is the relative velocity, are an exception~\cite{Dvorkin:2013cea,Slatyer:2018aqg,Boddy:2018wzy,Becker:2020hzj,Buen-Abad:2021mvc,Nguyen:2021cnb,Boddy:2022tyt,He:2025npy} (see also Refs.~\cite{Tashiro:2014tsa,Munoz:2015bca,Das:2025bnr} for analyses of other cosmological and astrophysical probes). The $v^{-4}$ scaling arises, for instance, in some freeze-in models~\cite{Dvorkin:2019zdi,Dvorkin:2020xga}, where the initial DM velocity distribution is non-thermal.

Remarkably, the collisionless nature of DM can be tested to high accuracy using cosmological data even within a completely secluded dark sector. This is the case if, for example, DM scatters with other relativistic species usually referred to as Dark Radiation~(DR)~\cite{Buen-Abad:2015ova,Lesgourgues:2015wza,Cyr-Racine:2015ihg,Bringmann:2016ilk,Archidiacono:2017slj,Buen-Abad:2017gxg,Archidiacono:2019wdp,Rubira:2022xhb,Mazoun:2023kid,Bagherian:2024obh,Buen-Abad:2025bgd,Cvetko:2025kda}. Investigating this possibility is timely, given the negative results of searches for DM interactions with the SM along a broad experimental frontier~\cite{Cirelli:2024ssz}. 

In these models, the strength of DM-DR interactions depends on the ratio between the conformal momentum transfer rate of DM particles to DR, $\Gamma_{\chi\text{-}\mathrm{DR}}$, and the conformal Hubble parameter, $\mathcal{H}$. Strong coupling is realized if $\Gamma_{\chi\text{-}\mathrm{DR}}/\mathcal{H}\gg 1$, weak coupling corresponds to $\Gamma_{\chi\text{-}\mathrm{DR}}/\mathcal{H} \lesssim 1$, while DM is kinetically decoupled from DR if $\Gamma_{\chi\text{-}\mathrm{DR}}/\mathcal{H} \ll 1$. Since the momentum transfer rate depends on the microscopic Lagrangian of the model, it is convenient to adopt a phenomenological parametrization~\cite{Cyr-Racine:2015ihg},
\begin{align}\label{eq:Gamma_chi_DR_general}
    \Gamma_{\chi\text{-}\mathrm{DR}} \equiv \frac{4}{3} \omega_{\rm DR} a_{{\rm D}} (1+z)^{n+1}\,,
\end{align}
where $\omega_{\rm DR} \equiv \Omega_{\rm DR}h^2$. By momentum conservation, it then follows that the momentum transfer rate of DR to DM is (see e.g.~Ref.~\cite{Buen-Abad:2017gxg})
\begin{align}\label{eq:Gamma_DR_chi_general}
    \Gamma_{\mathrm{DR}\text{-}\chi} = \frac{3\bar{\rho}_\chi}{4\bar{\rho}_{\rm DR}}\, \Gamma_{\chi\text{-}\mathrm{DR}} = \omega_\chi a_{\rm D} (1+z)^n\,,
\end{align}
where $\bar{\rho}_\chi \propto (1+z)^{3}$ and $\bar{\rho}_{\mathrm{DR}}\propto (1+z)^{4}$ are the time-dependent background energy densities in DM and DR, and $\omega_\chi \equiv \Omega_\chi h^2$. For a given Lagrangian realization, the integer $n$ parametrizing the time dependence of the interaction and the interaction strength parameter $a_{\rm D}$ can be computed in terms of microscopic quantities.\footnote{In this paper we mostly follow the conventions of Ref.~\cite{Cyr-Racine:2015ihg}, except for the conformal DM and DR momentum transfer rates, which are here defined with the opposite sign: $\Gamma_{\chi\text{-}\mathrm{DR}} = -\, \dot{\kappa}_{\chi}$ and $\Gamma_{\mathrm{DR}\text{-}\chi} = -\, \dot{\kappa}_{\mathrm{DR}\text{-}\mathrm{DM}}$. This gives a simpler matching to the underlying microscopic description. We also define $\Omega_{i} \equiv \bar{\rho}_{i,\hspace{0.2mm} 0}/\rho_{\rm cr}$ ($i = \mathrm{DR},\chi$), with $\rho_{\rm cr} = 3 H_0^2 M_{\rm Pl}^2$ and $M_{\rm Pl}^2 \equiv (8\pi G_N)^{-1}$.} Notice that $\omega_\chi \gg \omega_{\rm DR}$ in any realistic scenario, which implies $\Gamma_{\mathrm{DR}\text{-}\chi} \gg  \Gamma_{\chi\text{-}\mathrm{DR}}$ at low redshift. Comparing to the expansion of the Universe, one finds
\begin{equation}
    \frac{\Gamma_{\chi\text{-}\mathrm{DR}}}{\mathcal{H}}\;{\propto}\,
    \begin{cases}
     (1+z)^{n} \quad & \rm RD\,, \\
     (1+z)^{n+1/2} \quad &\rm MD\,, \\
     (1+z)^{n+2} \quad &\Lambda \rm D\,,
\end{cases}
\label{eq:G/H}
\end{equation}
in the radiation-~(R), matter-~(M) and cosmological constant-~($\Lambda$) dominated eras. These scalings entail that for $n\ge0$ the interaction rate is more important in the early Universe, and viceversa for $n\leq -1$. Previous literature considered interactions of DM with DR characterized by $n\ge 0$. It was found that DM needs to be very weakly coupled by the time of recombination~\cite{Becker:2020hzj,Archidiacono:2019wdp,Rubira:2022xhb}, and the scaling with $n\geq 0$ then implies the scattering is completely negligible in the late Universe.

These results, however, leave open the possibility that DM is very weakly coupled to DR around recombination, but its interaction strength increases at late times, resulting in weak or even strong coupling at low redshift. We refer to such dynamics, which corresponds to $n\leq -\hspace{0.2mm}1$, as {\it Dark Matter Recoupling} because $\Gamma_{\chi\text{-}\mathrm{DR}}/\mathcal{H}$ increases as the redshift decreases, both in the radiation- and matter-dominated eras. The main goal of this paper is to provide the first comprehensive study of the cosmological signatures of DM recoupling, focusing on the constraints that recent CMB and LSS observations set on a model that realizes $n = -\hspace{0.2mm}1$.\footnote{Kinetic recoupling of DM has been considered before, but with quite different dynamics: either a temporary recoupling to DR due to a rapid change in amplitude of the momentum transfer rate~\cite{Lehmann:2023vjv}, or a phase of recoupling to non-relativistic SM baryons, whose number density is enhanced by the baryon-antibaryon asymmetry~\cite{Kamada:2016qjo}.} We will find that, if the totality of DM is interacting, then it has to be weakly coupled today even in this scenario, thus establishing the almost-collisionless nature of DM throughout all the scales that are currently accessible to cosmological observations. On the other hand, we will argue that present data cannot exclude the possibility that a $\sim 4\%$ fraction of DM is strongly coupled to DR today. This is quite relevant in light of the upcoming data releases of the Dark Energy Spectroscopic Instrument (DESI) \cite{DESI:2025fxa}, Euclid \cite{Euclid:2024yrr} and the Rubin Observatory \cite{LSSTDarkEnergyScience:2018jkl}, which will measure the distribution of galaxies at low redshift with unprecedented precision and could further improve upon the constraints presented here. 

The rest of the discussion is organized as follows. In Sec.~\ref{sec:DM_recoupling} we present a microscopic quantum field theory model that realizes DM recoupling with $n = -\hspace{0.2mm}1$, and perform the matching to the phenomenological parametrization in Eq.~\eqref{eq:Gamma_chi_DR_general}. Then, in Sec.~\ref{sec:cosmo} we discuss the cosmological dynamics in terms of the main macroscopic parameters $a_{\rm D}$ and $\Delta N_{\rm eff}$ (the latter being equivalent to $\omega_{\rm DR}$). We also present analytical solutions that will allow us to intuitively understand the physics of recoupling. Section~\ref{sec:data} reports constraints on the macroscopic model parameters, obtained from measurements of the CMB temperature and polarization power spectra \cite{Aghanim:2019ame,AtacamaCosmologyTelescope:2025blo}, as well as CMB lensing \cite{ACT:2023dou},  and Baryon Acoustic Oscillations (BAO) from DESI \cite{DESI:2025zgx}.
Section~\ref{sec:implications_model} discusses the implications of these constraints for the parameter space of the microscopic model, including the relation to DM production via thermal freeze out. Finally, Sec.~\ref{sec:outlook} summarizes our findings and lays out future directions. Four appendices (\ref{app:mom_transf_rate},~\ref{app:DR_selfscatt},~\ref{app:linear} and~\ref{app:Thermal_details}) complete the paper, providing further details on the supporting calculations.

\section{Dark Matter Recoupling}\label{sec:DM_recoupling}
DM recoupling scenarios are realized if the DM momentum transfer rate in Eq.~\eqref{eq:Gamma_chi_DR_general} is constant or increases with the scale factor, $n \leq -1$. The analogy with Compton scattering between SM electrons and photons~\cite{Ma:1995ey}, as well as previous studies in secluded dark sectors~\cite{Cyr-Racine:2015ihg,Bringmann:2016ilk}, indicate that scenarios where DM interacts with DR composed of massless spin-$1$ bosons always lead to $n \ge 0$, with the inequality saturated by dark gluons of a non-Abelian gauge symmetry~\cite{Buen-Abad:2015ova,Lesgourgues:2015wza,Cyr-Racine:2015ihg}.

We are thus led to consider DR in the form of a light spin-$0$ field $\phi$, coupled to a fermionic DM particle $\chi$ through a Yukawa interaction. In Minkowski space, the dark sector Lagrangian is
\begin{equation}
\mathcal{L} = -\frac{1}{2}\partial_{\mu}\phi\partial^{\mu}\phi -\frac{1}{2}m_{\phi}^2\phi^2 - \overline{\chi}\left(\gamma^{\mu}\partial_{\mu}+m_{\chi}\right)\chi - y_\chi \phi\hspace{0.2mm}\overline{\chi}\chi -\frac{g_\phi}{3!}\phi^3-\frac{\lambda_\phi}{4!}\phi^4\,,
    \label{eq:Lag}
\end{equation}
where we assumed the mostly-plus convention for the metric. Notice that the cubic scalar self-coupling $g_\phi$ cannot be consistently set to zero, as it would be anyway generated by radiative corrections. The naturalness of the model is assessed in Sec.~\ref{sec:implications_model}.

\begin{figure}[t]
    \centering
\includegraphics[width=0.9\linewidth]{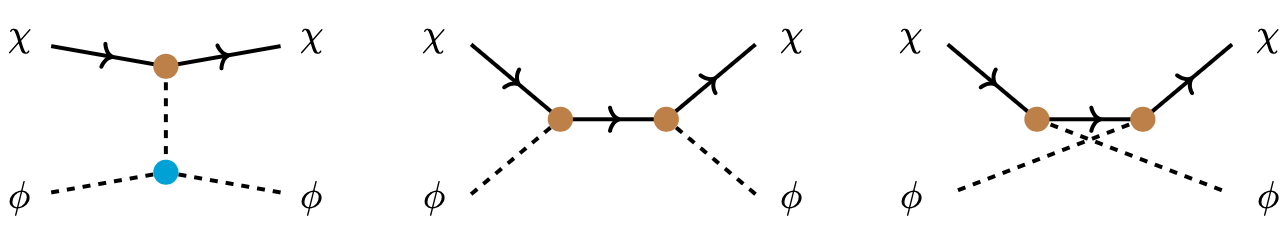}
    \caption{Feynman diagrams for elastic $\mathrm{DM}\text{-}\mathrm{DR}$ scattering at tree level. From left to right, the $t$-, $s$- and $u$-channel contributions. The {\color{cyan}\bf cyan vertex} corresponds to the cubic self-coupling {\color{cyan}$\bm{g_\phi}$} of the DR scalar $\phi$, whereas {\color{brown}\bf brown vertices} involve the Yukawa coupling {\color{brown}$\bm{y_\chi}$} between $\phi$ and the fermionic DM $\chi$.}
    \label{fig:feyn}
\end{figure}

The amplitude for the DM-DR elastic scattering process $\chi (p_1) \phi (p_2) \rightarrow \chi(p_3) \phi(p_4)$ is 
\begin{align}
    \mathcal{M}(\chi\phi \rightarrow \chi \phi) = \bar{u}(p_3) \left[ y_\chi^2 \left(\frac{i\slashed{p}_2 - 2 m_\chi}{s - m_\chi^2} + \frac{-i \slashed{p}_4 - 2 m_\chi}{u - m_\chi^2}\right) -  \frac{y_\chi g_\phi}{t - m_\phi^2}\right]u(p_1)\,,
    \label{eq:scattering amplitude}
\end{align}
corresponding to the three Feynman diagrams in Fig.~\ref{fig:feyn}. Initial-state~(final-state) momenta are taken ingoing~(outgoing). The DR momentum transfer rate is then given by~\cite{Cyr-Racine:2015ihg} (see Appendix~\ref{app:mom_transf_rate} for more details)
\begin{align}
\label{eq:Gamma DR-DM from ETHOS}
    \Gamma_{\mathrm{DR}\text{-}\chi} = -\, \frac{3 \bar{\rho}_\chi}{4 \bar{\rho}_{\mathrm{DR}}} \frac{(1+z)^{-1}}{16 \pi m_\chi^3} \frac{\eta_{\mathrm{DR}}}{3} \int_0^\infty \frac{\dd p\, p^2}{2 \pi^2} p^2 \frac{\partial f_{\mathrm{DR}}^{(0)}(p)}{\partial p} \big[A_0(p)-A_1(p) \big]\,,
\end{align}
where
\begin{align}
\label{eq: projection onto the lth Legendre polynomial}
    A_\ell (p) = \frac{1}{2}\int_{-1}^1 \dd \tilde{\mu} \,P_\ell(\tilde{\mu}) \frac{1}{\eta_{\chi}\eta_{\rm DR}}\Bigg(\sum_{\rm spins} |\mathcal{M}|^2\Bigg)_{\substack{s\; =\; \left[E_\chi(p) + E_{\mathrm{DR}} (p)\right]^2\\ \hspace{-10.75mm}t\; =\; 2 p^2(\tilde{\mu}-1)}}\,.
\end{align}
In the above expressions, $\eta_\chi = 2$ and $\eta_{\mathrm{DR}} = 1$ count the DM and DR degrees of freedom and $E_{\chi, \mathrm{DR}}(p) = \big(m_{\chi, \phi}^2 + p^2\big)^{1/2}$ are the particle energies, with $p$ and $\tilde{\mu}$ denoting the three-momentum and the cosine of the scattering angle in the center of mass frame. We denote with $f_{\mathrm{DR}}^{(0)}$ the background distribution function of DR, here assumed to take a Bose-Einstein form with temperature $T_{\rm DR}$, namely $f_{\rm DR}^{(0)}(p) = \big[\mathrm{exp} (E_\mathrm{DR}(p)/T_{\rm DR}) - 1\big]^{-1}$. Finally, $P_{\ell}$ indicates the Legendre polynomials. We can separate the contributions to the squared amplitude and momentum transfer rate as
\begin{align}
\sum_{\rm spins} |\mathcal{M}|^2 \equiv&\; \mathcal{M}^2_{g_\phi^2 y_\chi^2} +\mathcal{M}^2_{y_\chi^4} + \mathcal{M}^2_{g_\phi y_\chi^3}  \,, \label{eq: squared scattering amplitude}
\\
\Gamma_{\mathrm{DR}\text{-}\chi}  \equiv&\; \Gamma_{\mathrm{DR}\text{-}\chi\, (g_\phi^2 y_\chi^2)} +  \Gamma_{\mathrm{DR}\text{-}\chi\, (y_\chi^4)} +\Gamma_{\mathrm{DR}\text{-}\chi\, (g_\phi y_\chi^3)} \,, 
\label{eq: DR-DM momentum transfer rate}
\end{align}
in terms of the interaction in the $t$-channel alone, the $s/u$-channel terms alone, and their interference.

We now show that the $t$-channel exchange leads to DM recoupling to DR at late times. The spin-summed square of the scattering amplitude reads
\begin{align}\label{eq:M2_tchannel}
    \mathcal{M}_{g_\phi^2 y_\chi^2}^2 = 2\hspace{0.2mm} g_\phi^2 y_\chi^2 \frac{(4 m_\chi^2-t)}{(t-m_\phi^2)^2}\;,
\end{align}
which we insert in Eq.~\eqref{eq:Gamma DR-DM from ETHOS} to arrive, after integration by parts, at 
\begin{align}\label{eq:Gamma_DRchi_intermediate}
    \Gamma_{\mathrm{DR}\text{-}\chi\,(g_\phi^2 y_\chi^2)} \simeq \frac{ \omega_\chi}{\omega_{\mathrm{DR}}} \frac{(1+z)^{-2}}{128 \pi^3 m_\chi}\,  g_\phi^2 y_\chi^2 \left( \int_0^\infty \dd p\,   \frac{p^3 f_{\mathrm{DR}}^{(0)}(p)}{\big( p^2 + \tfrac{1}{4} m_\phi^2\big)^2} + \frac{\pi^2}{6} \frac{T_{\rm DR}^2}{m_\chi^2} \right)\,.% \left( 1 + \mathcal{O}\left(\frac{m_\phi}{T_{\rm DR}}\right)\right)\right]\,, %+ \mathcal{O}\bigg(\frac{1}{m_\chi^3}\bigg)
\end{align}
The second term in parentheses is very suppressed because the scatterings relevant for cosmological observables involve highly non-relativistic DM particles, which satisfy $T_{\rm DR} \ll m_\chi$; its expression neglects corrections proportional to $m_\phi/T_{\rm DR} \ll 1$, since by assumption $\phi$ behaves as radiation. However, the first term in parentheses diverges for $m_\phi \rightarrow 0$, as the scalar mass regulates both the collinear $(\tilde{\mu}\to 1)$ and soft ($p\to 0$) divergences of the momentum transfer rate in Eq.~\eqref{eq:Gamma DR-DM from ETHOS}. This prevents us from naively expanding the integrand in powers of $m_\phi^2/T_{\rm DR}^2$. To extract the asymptotic behavior at small $m_\phi/T_{\rm DR}$, we employ a technique~\cite{Rubira:2022xhb} inspired by the method of regions~\cite{Beneke:1997zp}, see Appendix~\ref{sec:regions} for details. We arrive at
\begin{align}\label{eq:Gamma_g2y2}
 \Gamma_{\mathrm{DR}\text{-}\chi\,(g_\phi^2 y_\chi^2)} \simeq  \frac{\omega_\chi}{\omega_{\rm DR}} \frac{(1+z)^{-1}}{128 \pi^3 m_\chi}\, g_\phi^2 y_\chi^2 \mathcal{C}_{g_\phi^2 y_\chi^2} \frac{T_{\rm DR,0}}{m_\phi} \left[1 + \mathcal{O} \left( \frac{m_\phi}{T_{\rm DR}}\log \frac{m_\phi}{T_{\rm DR}} \right)\right]\,, 
\end{align}
where $\mathcal{C}_{g_\phi^2 y_\chi^2} = \big[7 \sqrt{3} \log \left(2+ \sqrt{3}\right) - 6\big] \big/ 9 \approx 1.11$ and we have used the fact that the temperature of DR evolves as $T_{\rm DR} = T_{\rm DR,0}(1+z)$. Hence the momentum transfer rate $\Gamma_{\mathrm{DR}\text{-}\chi\,(g_\phi^2 y_\chi^2)}$ is proportional to $(1+z)^{-1}$, leading to DM recoupling at late times.

In the ETHOS parametrization~\cite{Cyr-Racine:2015ihg} introduced in Sec.~\ref{sec:intro}, Eq.~\eqref{eq:Gamma_g2y2} translates into
\begin{align}
\label{eq:ad}
    n = -\,1\,, \qquad a_{\rm D} =  \frac{g_\phi^2 y_\chi^2 \mathcal{C}_{g_\phi^2 y_\chi^2}}{128 \pi^3 m_\chi \omega_{\rm DR}}\frac{T_{\rm DR,0}}{m_\phi}\,.
\end{align}
Accordingly, the momentum transfer rate of DM to DR is independent of redshift,
\begin{align}
\Gamma_{\chi\text{-}\mathrm{DR}\, (g_\phi^2 y_\chi^2)} = \frac{4}{3}\hspace{0.2mm} \omega_{\rm DR} a_{\rm D}\,.  \label{eq:Gamma_chi_DR}
\end{align}
Notice that the usual ETHOS correspondence~\cite{Cyr-Racine:2015ihg} between the matrix element squared and the momentum transfer rate, $\sum_{\rm spins}|\mathcal{M}|^2 \propto p^{n-2} \leftrightarrow \Gamma_{\mathrm{DR}\text{-}\chi} \propto (1+z)^n$, fails in this case. The matrix element~\eqref{eq:M2_tchannel} for $m_\phi = 0$ would naively suggest $n = -\,2$, but the appearance of an additional factor $T_{\rm DR}/m_\phi$ in the momentum transfer rate, which regulates the divergence, results in $\Gamma_{\mathrm{DR}\text{-}\chi} \propto (1 + z)^{-1}$ instead. Incidentally, we note that the $n = -1$ scaling is obtained also for scalar DM coupled to scalar DR~\cite{Bringmann:2016ilk}. That model contains more free parameters (at the level of the renormalizable Lagrangian) and its analysis is left for future work.

The expression of the rate in Eq.~\eqref{eq:Gamma_g2y2} is valid at sufficiently low temperatures, $T_{\rm DR} < T_{\rm c}$ where $T_{\rm c}\sim m_\phi/\lambda_\phi^{1/2}$ is the critical temperature for the theory in Eq.~\eqref{eq:Lag}. Above $T_{\rm c}$, the infrared divergences are instead regulated by the thermal mass acquired by the scalar field, $m_{\widehat{\phi},T}^2 \sim \lambda_\phi\hspace{0.2mm} T_{\rm DR}^2$.\footnote{At $T_{\rm DR} > T_{\rm c}$, we evaluate the momentum transfer rate by modifying the DR dispersion relation from $E_{\rm DR}^2 = m_\phi^2 + p^2$ to $E_{\rm DR}^2 = m_{\widehat{\phi},T}^2 + p^2$ in both the matrix element and the distribution functions. Had we, instead, replaced $m_\phi \to m_{\widehat{\phi},T}$ only in the matrix element and taken a massless distribution function, $f_{\rm DR}^{(0)}(p) = [\exp(p/T_{\rm DR})-1]^{-1}$, then the first term in parentheses in Eq.~\eqref{eq:Gamma_DRchi_intermediate} would yield a nearly identical result: the numerical coefficient multiplying $m_{\widehat{\phi},T}/(16\hspace{0.3mm} T_{\rm DR})$ would be $\pi /2 \approx 1.57$ instead of $\mathcal{C}_{g_\phi^2 y_\chi^2} \approx 1.11$.} Furthermore, the scalar cubic self-coupling receive important thermal corrections that suppress its size as $g_{\widehat{\phi},T} \sim g_\phi\hspace{0.2mm} T_{\rm c}^2/T^2_{\rm DR}$  above the critical temperature. As a result, $\Gamma_{\mathrm{DR}\text{-}\chi\,(g_\phi^2 y_\chi^2)}$ is very quickly suppressed at $T_{\rm DR} > T_{\rm c}$, with redshift dependence corresponding to $n = -\,6$. This means that for the DM recoupling dynamics to be realized as discussed in this paper, the critical temperature needs to be sufficiently high. We will come back to these aspects in Sec.~\ref{sec:implications_model}.

We now turn to the other terms in the rate of Eq.~\eqref{eq: DR-DM momentum transfer rate}. The piece associated to the $s/u$-channel diagrams alone evaluates to
\begin{align}\label{eq:Gamma_DR_y4}
\Gamma_{\mathrm{DR}\text{-}\chi\,(y_\chi^4)} \simeq \frac{\omega_\chi}{\omega_{\rm DR}} \frac{(1+z)^{2}}{m_\chi^3} \frac{\pi y_\chi^4}{360} T_{\rm DR,0}^4 \bigg[1 + \mathcal{O} \bigg( \frac{m_\phi^2}{T_{\rm DR}^2}\bigg) + \mathcal{O} \left( \frac{T_{\rm DR}}{m_\chi}\right)  \bigg]\,,
\end{align}
up to the negligible corrections within square parentheses. This corresponds to $n = 2$, hence, this term is important in the early Universe, where it initially keeps DM and DR in thermal equilibrium, but it is negligible for the late-time evolution probed by CMB and LSS observations. The piece corresponding to the interference between $t$- and $s/u$-channel diagrams is found to be
\begin{align}\label{eq:Gamma_gy3}
\Gamma_{\mathrm{DR}\text{-}\chi\,(g y^3)} \simeq  \frac{\omega_\chi}{\omega_{\rm DR}} \frac{(1+z)^{-1}}{m_\chi^2} \frac{g_\phi y_\chi^3 }{96\pi^3} \mathcal{C}_{g_\phi^2 y_\chi^2} m_\phi T_{\rm DR,0} - \frac{\omega_\chi}{\omega_{\rm DR}} \frac{1+z}{m_\chi^3} \frac{\zeta(3)g_\phi y_\chi^3}{8\pi^3} T_{\rm DR,0}^3\,,
\end{align}
where the first term required another application of the method of regions, see Appendix~\ref{sec:regions}. At temperatures below $T_{\rm c}$, the first term gives a subleading contribution to the $n = - 1$ scaling in Eq.~\eqref{eq:Gamma_g2y2}, with relative suppression by a factor $\sim (y_\chi m_\phi / g_\phi) (m_\phi/m_\chi )$ which is always $\ll 1$ in the parameter region where $g_\phi$ is large enough to be accessible to cosmological observations, due to the tiny mass ratio $m_\phi/m_\chi$. The second term in Eq.~\eqref{eq:Gamma_gy3} scales as $n = 1$ at low temperatures and has size $\sim (y_\chi m_\phi / g_\phi) (T_{\rm DR}^2/m_\chi^2)$ relative to Eq.~\eqref{eq:Gamma_g2y2}, which is $\ll 1$ because DM is very non-relativistic. In conclusion, the $\mathcal{O}(y_\chi^4)$ and $\mathcal{O}(g_\phi y_\chi^3)$ terms can be safely neglected in the late Universe. Henceforth, when writing $\Gamma_{\mathrm{DR}\text{-}\chi}$ and $\Gamma_{\chi\text{-}\mathrm{DR}}$ we always refer only to the dominant $\mathcal{O}(g_\phi^2 y_\chi^2)$ piece with recoupling time dependence ($n = -\,1$), unless specified otherwise.

To summarize, the following picture emerges for the dark sector evolution. Early on, the Compton-like scattering mediated by $y_\chi$ maintains chemical and kinetic equilibrium between DM and DR. In this phase, DM can acquire its observed abundance through the freeze out of $\chi\overline{\chi}\to \phi\phi$ annihilation (see Sec.~\ref{sec:implications_model}). Once Compton scattering becomes ineffective, an intermediate stage follows when the two species are fully decoupled from each other. Finally, the recoupling dynamics kicks in, which is the focus of this work.  

The last ingredient required for the evaluation of the scattering rate is the energy density of DR. This can be parametrized in terms of the ratio between the temperature of the dark sector and the photon temperature, $\xi \equiv T_{\rm DR}/T_\gamma$, which is itself a function of temperature. Since $\phi$ is a boson, we have
\begin{align}\label{eq:omegaDR_xi0}
    \omega_{\rm DR} =  \frac{\eta_{\rm DR}}{2} \xi_0^4 \hspace{0.3mm}\omega_\gamma \qquad (\text{bosonic DR})\,,
\end{align}
where $\xi_0$ is the temperature ratio today (and at CMB last scattering). As the value of $\xi_0$ depends on dynamics in the very early Universe, about which we wish to remain agnostic, we take it as an additional free parameter of the model. Alternatively, we could define the temperature of DR in terms of its contribution to the effective relativistic degrees of freedom,
\begin{align} \label{eq:oDR_Neff}
        \omega_{\rm DR} = \frac{7}{8} \left( \frac{4}{11}\right)^{4/3}\,  \Delta N_{\rm eff}\,  \omega_{\gamma}\,,
\end{align}
such that, e.g.,~$\xi_0 = 0.5$ corresponds to $\Delta N_{\rm eff} \approx 0.14$.

In the discussion of the cosmological dynamics in Sec.~\ref{sec:cosmo}, we will use the macroscopic quantities $a_{\rm D}$ and $\Delta N_{\rm eff}$. These can be related to the microscopic Lagrangian parameters in Eq.~\eqref{eq:Lag} and the temperature ratio $\xi_0$ as
\begin{equation} \label{eq:macro_micro}
a_{\rm D} =  \frac{g_\phi^2 y_\chi^2 \mathcal{C}_{g_\phi^2 y_\chi^2}}{128 \pi^3 m_\chi m_\phi \hspace{0.2mm}\xi_0^3}\hspace{0.2mm} \frac{ 90 M_{\rm Pl}^2 H_0^2 }{\pi^2 T_{\gamma,0}^3 h^2} \,, \qquad \Delta N_{\rm eff} = \frac{4}{7}\left( \frac{11}{4}\right)^{4/3} \xi_0^4 \,,
\end{equation}
where $T_{\gamma,0} \approx 2.35 \times 10^{-4}\,\mathrm{eV}$ (or $\omega_\gamma = 2.47 \times 10^{-5}$). Combining Eqs.~\eqref{eq:Gamma_chi_DR},~\eqref{eq:oDR_Neff} and~\eqref{eq:macro_micro}, the DM momentum transfer rate normalized to the Hubble constant today is
\begin{align}
    \frac{\Gamma_{\chi\text{-}\mathrm{DR}}}{H_0} \approx 0.30 \left(\frac{0.68}{h}\right) \bigg(\frac{\xi_0}{0.5}\bigg) \bigg(\frac{m_\phi}{10^{-5}\;\mathrm{eV}}\bigg)\bigg(\frac{g_\phi/m_\phi}{2\times 10^{-5}}\bigg)^2 \bigg(\frac{y_\chi}{0.05}\bigg)^2 \bigg(\cfrac{1\;\mathrm{GeV}\,}{m_\chi}\bigg)\,,
\end{align}
where we have taken $m_\phi = 10^{-5}\;\mathrm{eV}$ and $\xi_0 = 0.5$ as reference parameters for DR. The reference Yukawa coupling $y_\chi$ and DM mass $m_\chi$, and in particular their combination $y_\chi^2/m_\chi$, are chosen to yield the observed abundance of DM via freeze out of $\chi \overline{\chi}\to \phi \phi$ annihilation in the early Universe, see Sec.~\ref{sec:implications_model}. As it will be quantified in Sec.~\ref{sec:data}, cosmological data require weak coupling today, $\Gamma_{\chi\text{-}\mathrm{DR}}/H_0 \lesssim 1$, thus imposing a non-trivial upper bound on the quantity $g_\phi/m_\phi$, which has dimension of a coupling constant.

Notice that the quartic self-coupling of the scalar, $\lambda_\phi$, does not enter the mapping in Eq.~\eqref{eq:macro_micro} above. Nevertheless, $\lambda_\phi$ does play a role in the cosmological evolution, for two reasons. First, it is primarily responsible for keeping DR always strongly coupled with itself (our calculation of the DR self-scattering rate is postponed to Appendix~\ref{app:DR_selfscatt}, together with a comparison to previous literature). Second, as already mentioned and as further discussed in Sec.~\ref{sec:implications_model}, $\lambda_\phi$ sets the critical temperature $T_{\rm c}$, above which the redshift evolution of the DM-DR momentum transfer rate changes dramatically.

\section{Cosmological Dynamics}\label{sec:cosmo}
At the level of the background, the first obvious difference with a standard cosmological scenario is the presence of DR. For the figures presented in this section, we assume the benchmark value $\Delta N_{\rm eff} = 0.1$.\footnote{Our fiducial cosmological model has $\omega_\chi = 0.120$, $\omega_b = 0.022$, $h=0.678$, $n_{\rm s} = 0.966$ and $A_{\rm s} = 2.1 \times 10^{-9}$.} The additional radiation delays matter-radiation equality and changes the comoving and angular diameter distances, with well known consequences for multiple cosmological observables (see Ref.~\cite{Saravanan:2025cyi} for a recent review). Notice that in our scenario DR behaves as a fluid, because the scalar quartic coupling $\lambda_\phi$ in Eq.~\eqref{eq:Lag} always keeps $\phi$ strongly self-coupled.

\begin{figure}
    \centering
\includegraphics[width=0.75\linewidth]{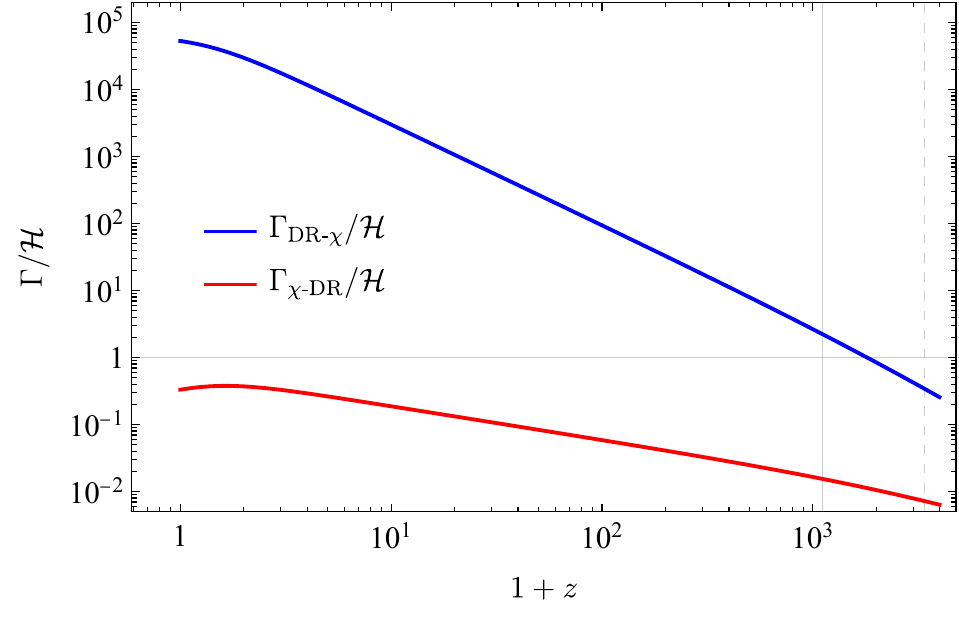}
    \caption{Ratio between the momentum transfer rates of DR~(blue) and DM~(red) and the conformal Hubble parameter $\mathcal{H}$. We set $\Delta N_{\rm eff}= 0.1$ and $a_{\rm D} = 100\;\mathrm{Mpc}^{-1}$. The vertical gray lines correspond to matter-radiation equality (dashed) and recombination (solid).}
    \label{fig:Gamma}
\end{figure}

Any other cosmological signature is controlled by the ratio $\Gamma_{\chi\text{-}\mathrm{DR}}/\mathcal{H}$. In our setup with $n = -\,1$, the DM momentum transfer rate is independent of redshift. For the Hubble rate, neglecting the effects of the cosmological constant and neutrino masses yields a simple approximate expression valid both in RD and MD, $\mathcal{H}/H_0 \simeq (\omega_r)^{1/2} (1+y)^{1/2}/(h\hspace{0.2mm} a_{\rm eq}\hspace{0.2mm} y)$, where $y \equiv a/a_{\rm eq}$ with $a_{\rm eq}$ being the scale factor at matter-radiation equality, and $\omega_r$ is the radiation density today (including the neutrinos, approximated as massless). Thus we arrive at 
\begin{align}
\label{eq:Gamma_chi}
\frac{\Gamma_{\chi\text{-}\mathrm{DR}}}{\mathcal{H}}  \simeq   \frac{\epsilon\hspace{0.2mm}y}{(1+y)^{1/2}}\,,\qquad \frac{\Gamma_{\mathrm{DR}\text{-}\chi}}{\mathcal{H}}  \simeq \frac{3\hspace{0.2mm} f_{\chi}}{4} \frac{\omega_r}{\omega_{\rm DR}}  \frac{\epsilon\hspace{0.2mm}y^2}{(1+y)^{1/2}}\,,   %\frac{4}{3} \frac{\omega_{DR}}{\sqrt{\omega_R}} \frac{a_D}{\tilde{H}_0} (1+z_{\rm eq})^{-1} \frac{y}{(1+y)^{1/2}} \equiv \frac{\Gamma_0}{\tilde{H}_0}\frac{\omega_R^{1/2}}{\omega_m} \frac{y}{(1+y)^{1/2}}\,,
\end{align}
where $f_\chi \equiv \omega_\chi / \omega_{\rm m}$ with $\omega_{\rm m} \equiv \omega_\chi + \omega_b$, and we defined the constant\footnote{The given numerical value of $\epsilon$ ignores a negligible contribution of $\Delta N_{\rm eff}$ to the total radiation energy density.}
\begin{equation}\label{eq:eps_def}
\epsilon \equiv \frac{a_{\rm eq} h}{(\omega_r)^{1/2}}\frac{\Gamma_{\chi\text{-}\mathrm{DR}} }{H_0} \approx 0.031 \left(\frac{h}{0.68}\right) \frac{\Gamma_{\chi\text{-}\mathrm{DR}} }{H_0}\,.
\end{equation}
In terms of the macroscopic parameters $\Delta N_{\rm eff}$ and $a_{\rm D}$, one has 
\begin{align}
   \frac{\Gamma_{\chi\text{-}\mathrm{DR}}}{H_0} \approx 0.33 \left(\frac{0.68}{h}\right)\left(\frac{\Delta N_{\rm eff}}{0.1}\right) \left(\frac{a_{\rm D}}{100 \,\text{Mpc}^{-1}}\right)\,.
\end{align}
The ratio between the momentum transfer rates and Hubble is shown in Fig.~\ref{fig:Gamma} assuming $a_{\rm D} = 100\;\mathrm{Mpc}^{-1}$, corresponding to $\epsilon \approx 1.0\times 10^{-2}$. We see that, for typical values of the cosmological parameters and $a_{\rm D} \lesssim  100\;\mathrm{Mpc}^{-1}$, DM is always weakly coupled to DR. Conversely, DR is much more strongly coupled to DM, with $\Gamma_{\mathrm{DR}\text{-}\chi}/\mathcal{H}\gtrsim1$ already at recombination. 
 
\subsection{Dark matter sound speed}\label{sec:DM_soundspeed}
The scattering of DR off DM particles heats up the latter, which in turn can affect structure formation if DM develops a sizable pressure perturbation, i.e.~a sizable sound speed (see Eq.~\eqref{eq:thetachi} below). In our recoupling scenario, DM is initially kept in thermal equilibrium with DR by the Compton-like $s$/$u$-channel interactions in Eq.~\eqref{eq:Gamma_DR_y4}. These decouple early on, too early to leave any signatures on the large-scale distribution of CMB temperature anisotropies or galaxies at late times. After the Compton-like scatterings decouple at $a_{\rm kd}$, the DR temperature always scales as $T_{\rm DR} \propto a^{-1}$, as its unperturbed distribution function preserves the Bose-Einstein thermal form. 

\begin{figure}
    \centering
\includegraphics[width=0.495\linewidth]{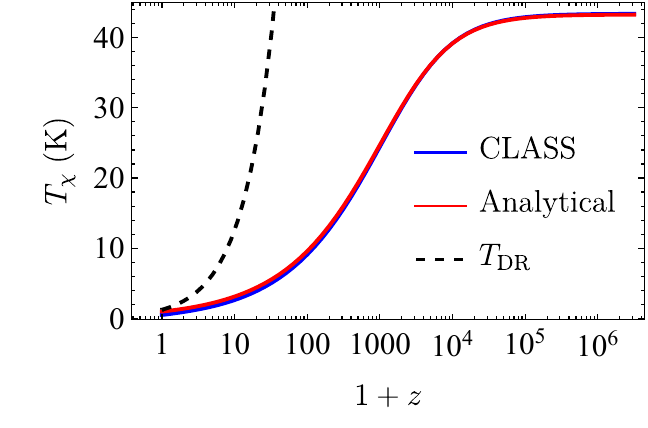}
\includegraphics[width=0.495\linewidth]{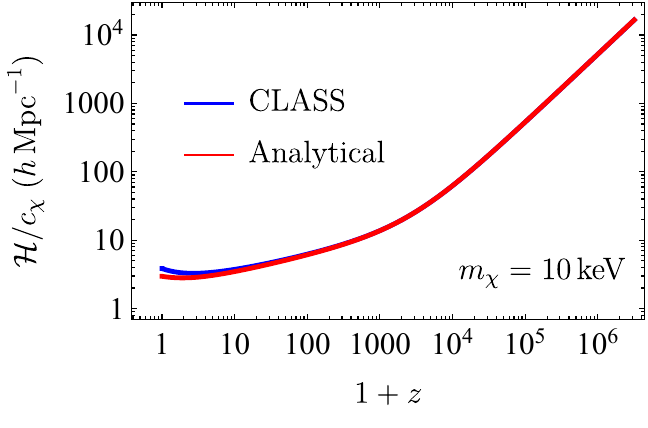}
    \caption{Evolution of the DM temperature and sound speed with redshift, for $\Delta N_{\rm eff}= 0.1$ and $a_{\rm D} = 100\;\mathrm{Mpc}^{-1}$. {\it Left:} DM temperature $T_\chi$, in Kelvin. For reference, we also show the DR temperature (black dashed curve). {\it Right:} Ratio $\mathcal{H}/c_\chi$, which approximately sets the DM Jeans wavenumber $k_{s,\chi}$ during matter domination, assuming $m_\chi = 10$~keV. DM perturbations are suppressed by thermal motions for $k\gtrsim k_{s,\chi}$, which however lies beyond the scales of relevance for this work.}
    \label{fig:TDM}
\end{figure}

On the other hand, the equation governing the evolution of the DM temperature, $T_\chi$, is found by integrating the unperturbed Boltzmann equation~(see e.g.~Refs.~\cite{Bringmann:2016ilk,Bringmann:2006mu,Bertschinger:2006nq}),
\begin{align} \label{eq:DM_T}
    \dot{T}_\chi + 2 \mathcal{H}T_\chi - 2\hspace{0.2mm}\Gamma_{\chi\text{-}\mathrm{DR}}\,(T_{\rm DR}-T_\chi) = 0\,,
\end{align}
where the dot indicates a derivative with respect to conformal time. The piece of~Eq.~\eqref{eq:DM_T} arising from DM-DR scattering depends on the DM momentum transfer rate. After $a_{\rm kd}$, only the $n = -\,1$ recoupling piece of $\Gamma_{\chi\text{-}\mathrm{DR}}$ is relevant. At first, the interaction is negligible and DM cools off adiabatically, $T_\chi\propto a^{-2}$. However, the interaction eventually recouples and the DM temperature is driven towards $T_{\rm DR}$. To obtain an analytic solution of Eq.~\eqref{eq:DM_T}, we find it convenient to switch again variables to $y = a/a_{\rm eq}$. Like we did above, we neglect $\Lambda$ and the neutrino masses, arriving at
\begin{align}\label{eq:Tchi_eq}
    T_\chi^{\hspace{0.1mm}\prime} + \frac{2}{y} T_\chi - \frac{2\hspace{0.2mm}\epsilon}{(1+y)^{1/2}} \left(\frac{T_{\rm DR,0}}{a_{\rm eq}y} - T_\chi\right) \simeq 0\,, %\frac{\Gamma_0 \Omega_R^{1/2}}{H_0 \Omega_m}\frac{(T_{DR}-T_\chi)}{\sqrt{1+y}} =0\,
\end{align}
where the prime denotes a derivative with respect to $y$. Its analytical solution is
\begin{align}\label{eq:Tchi_an}
    \frac{T_\chi (y)}{T_{\rm DR,0}} \simeq \frac{a_{\rm kd}}{a_{\rm eq}^2 y^2} + \frac{4 \epsilon}{3 a_{\rm eq}} \frac{\big[(1+y)^{1/2} - 1\big]^2\, \big[(1+y)^{1/2} + 2\big]}{y^2} + \mathcal{O}(\epsilon^2)\,,
\end{align}
taking $a_{\rm kd}$ as initial time for the evolution and expanding for $\epsilon \ll 1$. At early times during RD the interaction is still negligible, yielding the standard scaling $T_\chi \propto a^{-2}$. However, if at some point during RD the adiabatic and interaction terms in Eq.~\eqref{eq:Tchi_eq} become comparable, then DM rapidly transitions to a {\it constant}~temperature $T_\chi = \epsilon\hspace{0.3mm} T_{\rm DR, \hspace{0.1mm}eq}\,$, which lasts until matter-radiation equality. Afterwards, during MD, the DM temperature decreases again, as a result of the different time evolution of the Hubble parameter. The analytical expression in Eq.~\eqref{eq:Tchi_an} matches very well the full numerical solution obtained with \texttt{CLASS}, as shown in the left panel of Fig.~\ref{fig:TDM}. 

With Eq.~\eqref{eq:Tchi_an} in hand, we compute the (adiabatic) sound speed of DM: since the background pressure is $\overline{\mathcal{P}}_\chi = \bar{\rho}_\chi T_\chi / m_\chi$, we obtain
\begin{align}
    c_\chi^2 = \frac{\dot{\overline{\mathcal{P}}}_\chi}{\dot{\bar{\rho}}_\chi} = \frac{T_\chi}{m_\chi}\left(1-\frac{1}{3}\frac{d \log T_\chi}{d \log y}\right)\,,
\end{align}
whose evolution is shown in the right panel of Fig.~\ref{fig:TDM} for the smallest DM mass we consider in this paper, $m_\chi = 10$~keV (a discussion of the microscopic parameter space is found in Sec.~\ref{sec:implications_model}). During matter domination, $k_{s,\chi} = \sqrt{3/2}\,\mathcal{H}/c_\chi$ sets the characteristic wavenumber of DM above which perturbations are suppressed by thermal motions. We thus find that the sound speed becomes relevant at $k\gtrsim \mathrm{few}\,h\,\mathrm{Mpc}^{-1}$, beyond the scales relevant to this work. For larger DM mass the thermal suppression moves to even larger wavenumbers, as $k_{s,\chi} \propto m_\chi^{1/2}$. These results indicate that if DM is produced non-relativistically in the early Universe, then it can be treated as fully cold at all later times and for any choice of parameters, despite the recoupling dynamics.\footnote{The previous analysis applies if DM always remains weakly coupled. At strong coupling, $\Gamma_{\chi\text{-}\mathrm{DR}}/\mathcal{H}\gg 1$, the solution would be simply $T_\chi = T_{\rm DR} \propto a^{-1}$.}

\subsection{Linear perturbation theory}\label{sec:linear_theory}
At linear level, the continuity equation for DM is unchanged while the Euler equation is modified to take into account the momentum transfer to DR. In addition, we need the continuity and Euler equations describing the evolution of the DR perturbations. Following previous literature~\cite{Ma:1995ey,Cyr-Racine:2015ihg} we can write the system of equations, in the Newtonian gauge, as
\begin{align}
    \dot{\delta}_{\chi}+\theta_{\chi} -3 \dot{\phi} \,=&\; 0\,, \label{eq:deltachi}\\
    \dot{\theta}_{\chi} +\mathcal{H}\theta_{\chi} - c_\chi^2 k^2 \delta_\chi - k^2\psi \,=&\;  \Gamma_{\chi\text{-}\mathrm{DR}}(\tau)\,(\theta_{\rm DR} - \theta_{\chi})\,, \label{eq:thetachi} \\
   \dot{\delta}_{\rm DR} + \frac{4}{3}\theta_{\rm DR} -4 \dot{\phi} \,=&\; 0\,, \\
\label{eq:thetaDR}   \dot{\theta}_{\rm DR} - k^2 \left( \psi + \frac{\delta_{\rm DR}}{4}\right) \,=&\;  \Gamma_{\mathrm{DR}\text{-}\chi}(\tau)\, (\theta_{\chi} - \theta_{\rm DR} )\,, 
 \end{align}
where a sound speed term has been included for completeness in Eq.~\eqref{eq:thetachi}, but we henceforth set $c_\chi^2 = 0$ based on the results of Sec.~\ref{sec:DM_soundspeed}. Notice that we have made a fluid approximation for the DR species, retaining only the first two moments of the Boltzmann hierarchy. This is appropriate in our setup, because the quartic (and, to a lesser extent, the cubic) self-coupling in Eq.~\eqref{eq:Lag} keeps DR always strongly coupled with itself, allowing us to discard the shear and all the higher order moments. Further details and comparisons to previous work are provided in Appendix~\ref{app:DR_selfscatt}. The equations of motion for the perturbations of all the other species, namely baryons, photons and neutrinos, have the standard form. 

While the full set of equations is implemented in \texttt{CLASS}~\cite{Blas:2011rf} as described in Refs.~\cite{Archidiacono:2019wdp,Becker:2020hzj}, and solved without any approximations, it is nonetheless instructive to find analytical solutions where possible, to gain an intuition of the relevant magnitude and time-dependence of the new signatures introduced by DM interacting with DR in the late Universe.
In the simplest recoupling scenario, the evolution of the perturbations in the very early Universe is unchanged, which implies the initial conditions for the equations above and for Boltzmann codes in general can be straightforwardly implemented. In particular, in this work we will be concerned only with adiabatic initial conditions. Analogously, the evolution of the perturbations is not affected by the new interaction on scales much larger than the horizon, at least if DM is weakly coupled, which implies new physical effects will be located at late times and deeply inside the horizon. 
In this regime we can neglect any terms in the equations of motion involving time derivatives of the gravitational potentials, as they are strongly suppressed by powers of $\mathcal{H}/k \ll 1$. Even in this simplified setup, and further assuming matter domination in the background, an analytical solution to Eqs.~\eqref{eq:deltachi}$\,$-$\,$\eqref{eq:thetaDR} does not exist, as in practice we need to deal with a three-fluid system where the baryons also contribute to the evolution of the metric through the Poisson equation, $k^2\psi = -\,4 \pi G_N a^2 (\bar{\rho}_\chi \delta_\chi + \bar{\rho}_b \delta_b)$.
Progress can be made by looking for perturbative solutions at small $\Gamma_{\chi\text{-}\mathrm{DR}}/\mathcal{H}$ and correspondingly large $\Gamma_{\mathrm{DR}\text{-}\chi}/\mathcal{H}$, as illustrated by Fig.~\ref{fig:Gamma}. Let us first rewrite the equation of motion for the DR density perturbation as
\begin{align}\label{eq:deltaDR_eom}
    \delta_{\rm DR}'' + \left(\frac{1}{2} + \frac{\Gamma_{\mathrm{DR}\text{-}\chi}}{\mathcal{H}}\right) \frac{\delta_{\rm DR}'}{y}+
    \frac{k^2}{3 \mathcal{H}^2 y^2}\, \delta_{\rm DR}-\left(\frac{2}{y^2}\delta_m + \frac{4 \Gamma_{\mathrm{DR}\text{-}\chi}}{3\mathcal{H}y}\delta_\chi'\right)=0\,,
\end{align}
where matter domination has been assumed, with $\mathcal{H}\propto y^{-1/2}$. Here $\delta_m \equiv f_\chi \delta_\chi + (1-f_\chi)\delta_b$ is the total matter density fluctuation, and we recall that $f_\chi = \bar{\rho}_\chi/(\bar{\rho}_\chi + \bar{\rho}_b)$. If $\Gamma_{\mathrm{DR}\text{-}\chi}/\mathcal{H}\gg1$, as suggested by Fig.~\ref{fig:Gamma}, then in both parentheses in Eq.~\eqref{eq:deltaDR_eom} the first term can be safely neglected compared to the second one. Similarly we can discard the $\delta_{\rm DR}''$ term compared to $(\Gamma_{\mathrm{DR}\text{-}\chi}/\mathcal{H}) \delta_{\rm DR}'/y$. The equation of motion for DR thus becomes
\begin{align}
\label{eq:delta_DR_TC}
    \delta_{\rm DR}' + \frac{k^2}{3(\Gamma_{\mathrm{DR}\text{-}\chi}/\mathcal{H}) \mathcal{H}^2 y}\, \delta_{\rm DR} -\frac{4}{3}\delta_\chi' \simeq 0\,.
\end{align}
We can already draw two important conclusions. First, on large scales where the diffusion term above is not relevant, strong coupling makes the DR perturbation grow like the one of DM, $\delta_{\rm DR} \simeq 4\hspace{0.2mm} \delta_\chi/3$, where the $4/3$ prefactor matches the adiabatic initial condition for relativistic particles. Second, as modes enter the horizon they are not immediately washed away by radiation pressure, which would be the case if the DR perturbations propagated with a sound speed squared equal to $1/3$. Actually, the effective sound speed squared of the DR fluid is $1/(3\Gamma_{\mathrm{DR}\text{-}\chi}/\mathcal{H})\ll 1/3$, which delays the suppression of fluctuations once inside the horizon. In other words, the DR particles, which in the absence of interactions would naturally flee out of the potential wells generated by DM, are so strongly interacting that their perturbations are forced to grow and remain closer to DM overdensities. 

Equation~\eqref{eq:delta_DR_TC} is still coupled to the equations for the DM species, but in the latter DR enters as a perturbation suppressed by $\Gamma_{\chi\text{-}\mathrm{DR}}/\mathcal{H}$, which we assume to be small. Therefore, we make the further approximation of treating the source term in Eq.~\eqref{eq:delta_DR_TC} to zeroth order in the interaction. Notice, however, that even in the limit of zero momentum transfer the cosmological background differs from the one of a standard $\Lambda$CDM model, due to the presence of DR. Additional relativistic particles in the background leave a characteristic imprint in the transfer function of DM and baryons, as they further suppress structure growth during radiation domination and shift the value of the sound horizon. Since our primary interest is in the novel dynamics of recoupling, we will show the solutions of the equations of motion, and the accompanying figures, taking a $\Lambda$CDM~+~$\Delta N_{\rm eff}$ model (with free-streaming DR) as reference. For the latter we therefore have $\delta_\chi^{\Lambda\mathrm{CDM} + \Delta N_{\rm eff}} (k,y)\simeq \delta_m^{\Lambda\mathrm{CDM} + \Delta N_{\rm eff}} (k,y)\propto y$ in matter domination.

\begin{figure}
    \centering
    \includegraphics[width=0.475\linewidth]{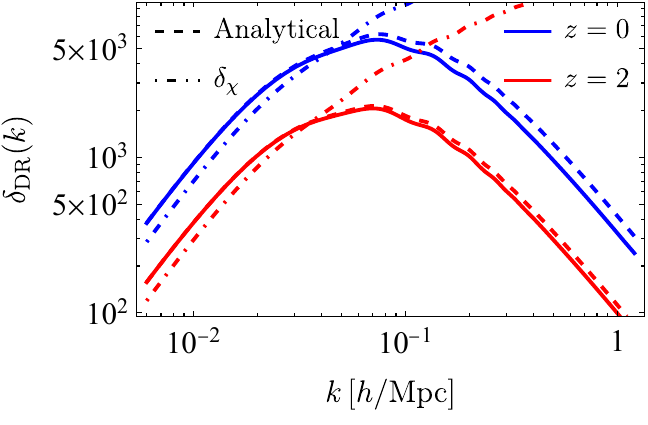}\hspace{2mm}
    \includegraphics[width=0.475\linewidth]{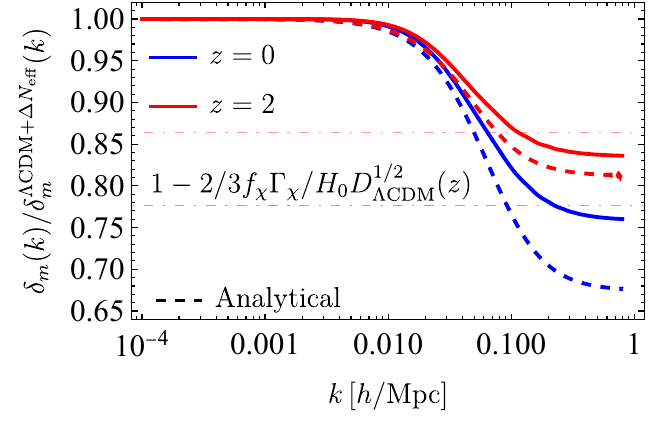}
    \caption{{\it Left:} The DR density perturbation $\delta_{\rm DR}$ as a function of wavenumber $k$, obtained from \texttt{CLASS} (solid) and our analytical approximation in Eq.~\eqref{eq:deltaDR_an}~(dashed). Also shown, for comparison, is the DM density perturbation $\delta_\chi$~(dot-dashed). {\it Right:} Ratio between the total matter density transfer function in a recoupling model and the corresponding quantity in a $\Lambda$CDM Universe with extra radiation. Solid~[dashed] lines correspond to \texttt{CLASS}~[our analytical solution in Eq.~\eqref{eq:deltam}]. Thin dot-dashed lines indicate the improved asymptotic ratio in Eq.~\eqref{eq:deltam_limit_impr}, which accounts for $\Lambda$ domination at low redshift. As in previous figures, we set $\Delta N_{\rm eff} = 0.1$ and $a_{\rm D} = 100$ Mpc$^{-1}$.}
    \label{fig:linear}
\end{figure}

The solution of Eq.~\eqref{eq:delta_DR_TC} under our set of assumptions reads then
\begin{align}\label{eq:deltaDR_an}
    \delta_{\rm DR}(k,y) \simeq \frac{4}{3} \hspace{-1mm} \left[1 - Y -  Y^2 e^{Y}\,
   \text{Ei}\left(-Y\right)
    \right] \delta_m^{\Lambda\mathrm{CDM} + \Delta N_{\rm eff}}(k,y)\,, \qquad \mathrm{Ei}\,(-Y) \equiv -\hspace{-1mm} \int_{Y}^{+\infty} \mathrm{d}t\hspace{0.4mm}\frac{e^{-t}}{t}\,, 
\end{align}
where $\mathrm{Ei}$ denotes the exponential integral function and
\begin{align}
\label{eq:ks}
Y(k, y) \equiv
  \frac{2 k^2}{k_{s,\mathrm{DR}}^2(y)} \,,
\quad k_{s,\mathrm{DR}}^2(y) \equiv 3\mathcal{H}^2 \left( \frac{\Gamma_{\mathrm{DR}\text{-}\chi}}{\mathcal{H}} \right) = k_{\ast,\mathrm{DR}}^2\, y^{1/2} \,, \quad k_{\ast,\mathrm{DR}}^2 \equiv \frac{9}{4} H_0^2 \frac{\omega_\chi}{\omega_{\rm DR}} \epsilon \,\Omega_{\rm m}\,.
\end{align}
The first factor in the definition of $k_{s,\mathrm{DR}}$ is the characteristic wavenumber for a non-interacting radiation species, whereas the factor in parentheses is due to the $\chi$-DR interaction and causes the suppression to move to larger $k$, as we assume $\Gamma_{\mathrm{DR}\text{-}\chi} \gg \mathcal{H}$. The effective characteristic wavenumber then scales as
\begin{align}\label{eq:ktilde_def}
   k_{s,\mathrm{DR}}(z) \approx \frac{0.067}{\mathrm{Mpc}} \left(\frac{f_\chi}{0.85}\right)^{1/2}\hspace{-1mm}\left(\frac{a_{\rm D}}{100\;\mathrm{Mpc}^{-1}}\right)^{1/2}\hspace{-0.5mm}(1+z)^{-1/4}\,.
\end{align}
The analytical expression in Eq.~\eqref{eq:deltaDR_an} has the correct limits,
\begin{align}\label{eq:deltaDR_limits}
    \delta_{\rm DR}(k,y)\hspace{-1mm} \overset{\strut k\, \ll\, k_{s,\mathrm{DR}}}{\longrightarrow} \hspace{-1mm} \frac{4}{3}\delta_m^{\Lambda\mathrm{CDM} + \Delta N_{\rm eff}}(k,y)\,,\quad \delta_{\rm DR}(k,y) \hspace{-1mm} \stackrel{\strut k\, \gg\, k_{s,\mathrm{DR}}}{\longrightarrow} \hspace{-1mm} \frac{4}{3} \frac{k_{s,\mathrm{DR}}^2(y)}{k^2}\,  \delta_m^{\Lambda\mathrm{CDM} + \Delta N_{\rm eff}}(k,y)\,,
\end{align}
recovering adiabatic growth at large scales and the typical $\propto k^{-2}$ suppression at small scales. The agreement between the analytical approximation and the numerical output of \texttt{CLASS} is shown in the left panel of Fig.~\ref{fig:linear}, confirming the accuracy of our solution.

The expression for the DR density perturbation in Eq.~\eqref{eq:deltaDR_an} should then be inserted in the equations of motion for DM. Unfortunately, the exponential integral function cannot be further integrated. A more analytically tractable approximation is  
\begin{align}
\label{eq:deltaDR_appr}
%\delta_{\rm DR}(k,y) = \frac{4}{3} \frac{\delta_m^{\Lambda\mathrm{CDM} + \Delta N_{\rm eff}}(k,y)}{1+ \tfrac{3}{2}\tilde{k}^2/ y^{1/2}} \,,   
\delta_{\rm DR}(k,y) \,\simeq\, \frac{4}{3} \frac{\delta_m^{\Lambda\mathrm{CDM} + \Delta N_{\rm eff}}(k,y)}{1 + \tfrac{3}{2} k^2/ k_{s,\mathrm{DR}}^2(y)}\,,
\end{align}
which is roughly 10\% accurate compared to the numerical solution. Equation~\eqref{eq:deltaDR_appr} exactly reproduces the low-$k$ analytical limit in Eq.~\eqref{eq:deltaDR_limits}, while the factor $3/2$ in the denominator is inserted to improve slightly the agreement with the numerical solution at high $k$ and very low redshift. We are now in a position to solve for the evolution of the DM and baryon fluctuations. The equation of motion for the total matter perturbation is, deep inside the horizon and during matter domination,
\begin{align}
\label{eq:deltam}
    \delta_m '' + \frac{3}{2y}\delta_m'-\frac{3}{2 y^2 }\delta_m-\frac{f_\chi\epsilon}{y^{1/2}}\left(\frac{3}{4}\delta_{\rm DR}'-\delta_\chi'\right)=0\,.
\end{align}
Since $\epsilon \ll 1$, we seek a solution of the form
\begin{align}
\label{eq:series}
    \delta_m = \delta_m^{\Lambda\mathrm{CDM}+\Delta N_{\rm eff}} + \epsilon\hspace{0.2mm}  \delta_m^{[1]} + \mathcal{O}(\epsilon^2)\,,
\end{align}
%where $\delta_m^{[0]}=\delta_\chi^{[0]}$ is the $\Lambda$CDM solution including the extra radiation in the background discussed before. 
where the lowest-order solution $\delta_m^{\Lambda\mathrm{CDM}+\Delta N_{\rm eff}}\simeq \delta_\chi^{\Lambda\mathrm{CDM}+\Delta N_{\rm eff}}$ grows linearly with the scale factor during matter domination. Plugging this ansatz into Eq.~\eqref{eq:deltam} and using Eq.~\eqref{eq:deltaDR_appr}, we obtain the analytical solution shown in the right panel of Fig.~\ref{fig:linear}. The explicit expression is not very illuminating and is presented for completeness in Appendix~\ref{app:linear}, together with additional results for the relative density perturbation between DM and baryons. The analytic approximation matches quite well with the \texttt{CLASS} results, and shows, as expected, a suppression with respect to a model without DM-DR interactions. Asymptotically, at small scales the matter transfer function is reduced by
\begin{align}
\label{eq:deltam_limit}\delta_m(k,y)\;\overset{\strut k\, \gg\, k_{s,\mathrm{DR}}}{\longrightarrow}\;\left(1-\frac{2}{3}f_\chi \epsilon\hspace{0.2mm} y^{1/2}\right) \delta_m^{\Lambda\mathrm{CDM}+\Delta N_{\rm eff}}(k,y)\,,
\end{align}
which shows how the new physics of recoupling is localized at very late times and grows rapidly with redshift. For reference, if $\epsilon = 10^{-2}$, the maximum suppression of the transfer function at $z=1$ is approximately 20\%. An improved analytical expression for the asymptotic ratio in Eq.~\eqref{eq:deltam_limit} can be found by replacing $y$ with the linear growth factor in a standard cosmology, which accounts for $\Lambda$ domination at very low redshift. We obtain
\begin{equation}
\label{eq:deltam_limit_impr}\delta_m(k,z)\;\overset{\strut k\, \gg\, k_{s,\mathrm{DR}}}{\longrightarrow}\; \bigg( 1 - \frac{2}{3} f_\chi \frac{ \Gamma_{\chi\text{-}{\rm DR}} }{H_0} D^{1/2}_{\Lambda\mathrm{CDM}}(z) \bigg)\delta_m^{\Lambda\mathrm{CDM}+\Delta N_{\rm eff}}(k,z)\,,
\end{equation}
where the growth factor is normalized to $1$ at $z = 0$. This improved asymptotic limit is shown by the thin dot-dashed lines in the right panel of Fig.~\ref{fig:linear}.

As Eq.~\eqref{eq:deltam_limit} suggests, we can trade the strength of the interaction for the fraction of DM coupled to DR, as long as $f_\chi \epsilon y^{1/2}\lesssim1$. We then conclude this discussion with a few words about the strong coupling scenario, $\Gamma_{\chi\text{-}\mathrm{DR}}/\mathcal{H}\gg 1$, for a small fraction of interacting DM, $f_\chi \ll 1$. The latter would be tightly coupled to DR, and therefore its perturbations would be prevented from growing on small scales. This is reminiscent of non-relativistic particles with a large sound speed, e.g.~massive neutrinos and other light relics, which do not cluster below some characteristic scale, also affecting the growth of cold DM and baryons~\cite{Lesgourgues:2006nd,Verdiani:2025jcf}. It is important to note that even if the two fluids are tightly coupled, no dark acoustic oscillations are produced. This is due to $\rho_{\chi}/\rho_{\rm DR} \gg 1$, where $\rho_\chi$ is the energy density in interacting DM, which causes any oscillations to be completely damped by friction.

\subsection{CMB and matter power spectra}
\begin{figure}[!tb]
    \centering
    \includegraphics[width=0.45\linewidth]{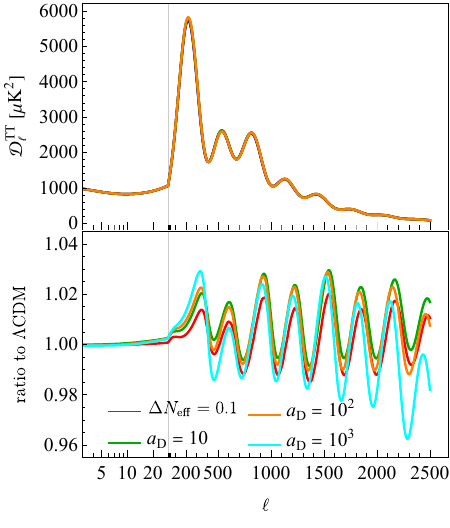}\hspace{3mm}
    \includegraphics[width=0.45\linewidth]{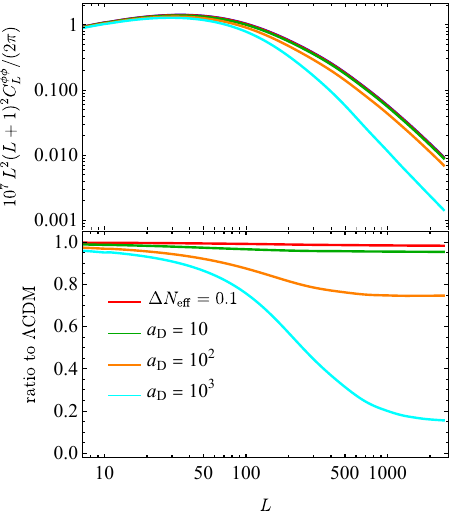}
    \caption{{\it Top:} Lensed TT CMB power spectrum (left) and CMB lensing power spectrum (right). A reference model with non-interacting free-streaming DR, assuming an energy density corresponding to $\Delta N_{\rm eff} = 0.1$, is shown in red. Other colors show increasing values of $a_{\rm D}$ (in $\mathrm{Mpc}^{-1}$). {\it Bottom:} Ratio to the corresponding $\Lambda$CDM Universe.}
    \label{fig:CMB_lensed}
\end{figure}

The discussion of the evolution of density perturbations presented above is useful to qualitatively understand the new features introduced by DM recoupling on the CMB temperature power spectrum and matter power spectrum, which are obtained from~\texttt{CLASS}. We begin with the former, shown in the top-left panel in Fig.~\ref{fig:CMB_lensed}. Different colors display different values of $a_{\rm D}$, with the $\Lambda$CDM power spectrum in blue, and a reference model with decoupled {\it free-streaming} DR in red.
In a recoupling scenario distances are the same as in a $\Lambda$CDM Universe with non-zero $\Delta N_{\rm eff}$, so the comoving distance to the last scattering surface is unchanged compared to the latter. Similarly, the physics of recombination is unaffected by the recoupling interaction, which is relevant only during the late matter-dominated era. Since we assume that DR is always self-interacting if $a_{\rm D}\ne 0$, its fluctuations propagate with a sound speed $c_s = 1/\sqrt{3}$ and undergo oscillations like the photon-baryon plasma, enhancing the amplitude of the oscillations in the CMB power spectrum compared to a $\Lambda$CDM Universe, and contrary to a $\Lambda$CDM model plus free-streaming DR.  As we are showing the lensed CMB power spectrum, the larger suppression of the matter clustering at low-$z$ with increasing values of $a_{\rm D}$ causes smearing and damping of the acoustic peaks at high $\ell$. This is the most visible signature of recoupling in the CMB power spectra, as one can see from the bottom-left panel in Fig.~\ref{fig:CMB_lensed}. Given current experimental uncertainties, we conclude that primary CMB data alone will not be able to constrain the product $\Delta N_{\rm eff} \times (a_{\rm D}/\mathrm{Mpc}^{-1})$ much better than roughly one hundred. Fortunately, CMB maps also allow for the extraction of the lensing power spectrum. This is shown in the right panels of Fig.~\ref{fig:CMB_lensed}, with the same color coding as above. Looking at the ratio to a $\Lambda$CDM model, we see that increasing the momentum transfer rate of DM particles leads to a stronger suppression of the lensing power spectrum, in line with the discussion in Sec.~\ref{sec:cosmo} on linear perturbation theory.
Large values of $a_{\rm D} \gtrsim 10^3\; \mathrm{Mpc}^{-1}$, corresponding to DM being strongly coupled to DR today, produce a $\gtrsim
80\%$ suppression of the lensing power, well above the current uncertainties measured, for example, by the Atacama Cosmology Telescope (ACT)~\cite{ACT:2023dou}.

\begin{figure}[!htb]
    \centering
    \includegraphics[width=0.475\linewidth]{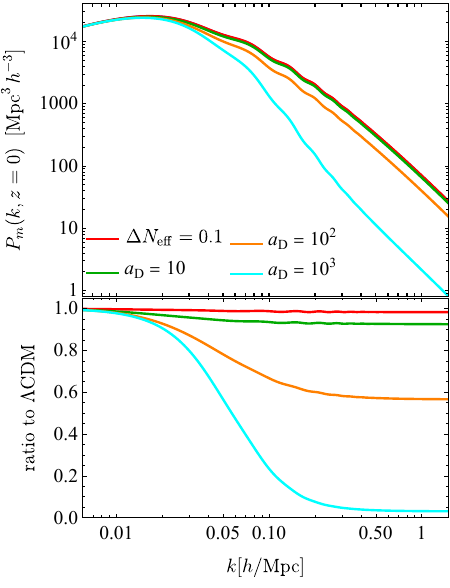}
     \includegraphics[width=0.475\linewidth]{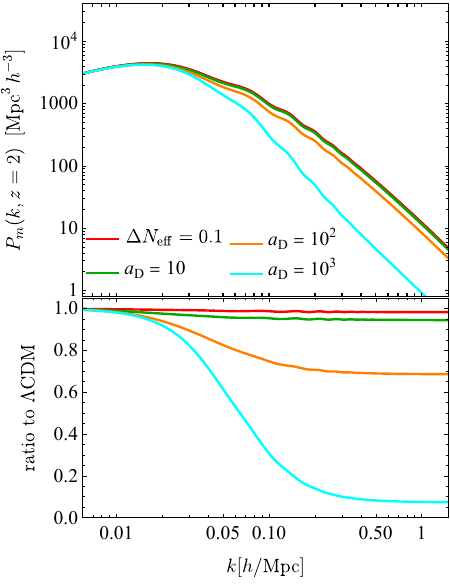}
    \caption{Same as in Fig.~\ref{fig:CMB_lensed}, but for the linear matter power spectrum $P_m(k,z)$ at redshift $z=0$ (left panel) and $z=2$ (right panel).}
    \label{fig:Pk}
\end{figure}

The linear matter power spectrum at $z=0$ and $z=2$ is shown in Fig.~\ref{fig:Pk} (left and right panel, respectively), using the same color coding as in Fig.~\ref{fig:CMB_lensed}. First, we observe the well-known result that extra free-streaming radiation, shown in red, causes a small suppression of power compared to a $\Lambda$CDM cosmology. As discussed in Sec.~\ref{sec:cosmo}, this is due to a longer phase of radiation domination, which prevents modes entering the horizon during this era to significantly grow. As we turn on DM recoupling, the power spectrum is further and further suppressed, as we have shown in Eq.~\eqref{eq:deltam_limit}, to the point that for strong coupling $\Delta N_{\rm eff} \times a_{\rm D} \sim 10^2\;\mathrm{Mpc}^{-1}$ there is almost no power left beyond $k\gtrsim 0.2$ $h$/Mpc, already at $z\sim 2$. This contrasts with the CMB power spectrum discussed above, which was basically insensitive to the new interaction for similar values of $\Delta N_{\rm eff}\times a_{\rm D}$. Our analysis therefore suggests that the galaxy power spectrum measured by DESI and Euclid could provide an invaluable test of whether DM scatters with other species in the late Universe. Enabling that test, however, requires the development of perturbation theory beyond the linear regime studied in this work, an issue we intend to return to in a forthcoming publication.

\section{Cosmological Constraints}\label{sec:data}
After a comprehensive discussion of the phenomenology of recoupling, in this section we finally present the constraints on the model parameters. These are obtained by fitting the model to CMB and BAO data, using the following definitions for the different datasets:
\begin{itemize}
    \item CMB: \textit{Planck 2018} likelihoods \textsc{COMMANDER} and \textsc{SimAll} for low-$\ell$ temperature and polarization, respectively, together with the \texttt{NPIPE} \textsc{CamSpec} likelihood for high-$\ell$ TT/EE/TE \cite{rosenberg22}. 
    CMB lensing is included through the ACT DR6 lensing likelihood, which combines Planck \texttt{NPIPE} and ACT measurements \cite{ACT:2023kun, ACT:2023dou, Carron:2022eyg}. The non-linear corrections to the matter power spectrum are obtained with the latest version of \texttt{HMCode} \cite{Mead:2020vgs}.\footnote{We are aware that these predictions might not be accurate for our model, which introduces late-time dynamics with a potentially large impact on structure growth. We will analyze the non-linear phenomenology in future work.} The CMB lensing power spectrum is fitted up to $L_{\rm max} = 763$ as per the ACT baseline.
    \item  CMB(P-ACT): Compared to the previous one, this combination also includes the recent ACT TT/EE/TE power spectra likelihoods \cite{AtacamaCosmologyTelescope:2025blo}, using ACT temperature (polarization) data for $\ell>1000$ ($\ell>600$) and Planck data for $\ell \le 1000$ ($\ell \le 600$).
    \item DESI BAO: Measurements of the transverse and line-of-sight BAO scale from DESI DR2 \cite{DESI:2025zgx}.
\end{itemize}
Parameter inference is performed using the MCMC sampler \cite{Lewis:2002ah,Lewis:2013hha} as implemented in \texttt{Cobaya} \cite{torrado:2020dgo}.
We assume flat priors on
the six standard cosmological parameters $\{ \omega_b, \omega_\chi, H_0, n_{\rm s}, A_{\rm s}, \tau \}$. For definiteness, the DM mass is fixed to $1$ GeV. However, the results presented in this section are insensitive to its value, as long as DM is highly non-relativistic by the time of recombination. Unless otherwise specified, we assume that the totality of DM is interacting with DR. The latter is modeled as a fluid of scalar particles.
Finally, we vary the two non-standard parameters $\Delta N_{\rm eff}$ and $\Delta N_{\rm eff} \times a_{\rm D}$ with top-hat priors $[0,1]$ and $[0,10^7]\;\mathrm{Mpc}^{-1}$, respectively. This parametrization is more efficient than sampling $\Delta N_{\rm eff}$ and $a_{\rm D}$ separately, as they are degenerate along a hyperbola: when $\Delta N_{\rm eff}$ is very small a very large interaction strength is allowed, while on the other branch of the hyperbola the constraint on $\Delta N_{\rm eff}$ is visible.\footnote{We explicitly tested the impact of separating the two parameters in the sampling.} The effective number of neutrinos is fixed to its standard cosmological value $N_{\rm eff}=3.044$, and the neutrino mass sum is set to the minimum value allowed by oscillations $\sum m_\nu = 0.06$ eV, unless otherwise specified.

\begin{figure}
    \centering
    \includegraphics[width=1\linewidth]{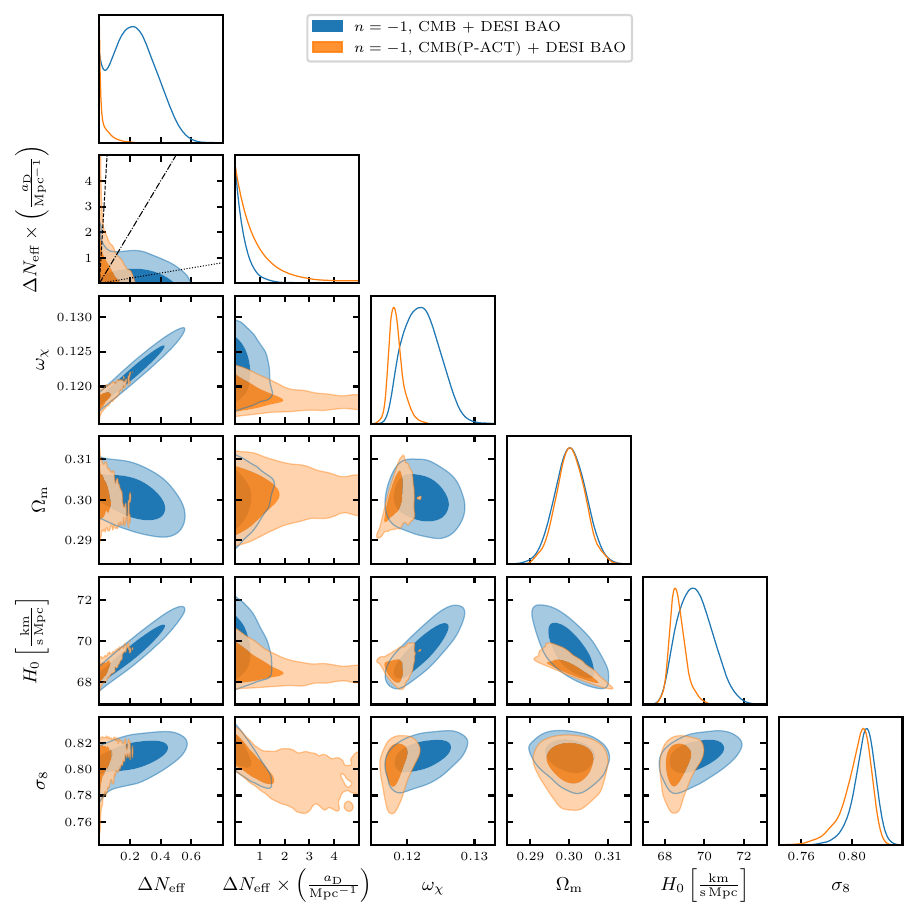}
    \caption{Marginalized 1D and 2D posteriors of the model parameters and of a selection of standard cosmological parameters. Constraints are obtained from CMB + DESI BAO {\color{lightblue}\bf(blue)} and CMB(P-ACT) + DESI BAO {\color{lightorange}\bf(orange)}. The thin black (dashed, dot-dashed, dotted) lines in the plane $\Delta N_{\rm eff} \times a_{\rm D}$ vs.~$\Delta N_{\rm eff}$ are drawn at constant $a_{\rm D}=(100,\,10,\,1)\;\mathrm{Mpc}^{-1}$.}
    \label{fig:lin_bounds}
\end{figure}

The constraints for the combination of  Planck + CMB lensing + DESI BAO data are shown in blue in Fig.~\ref{fig:lin_bounds}. We do not display the constraints from CMB alone. As expected, the strength of the interaction $a_{\rm D}$ is practically unconstrained by the primary CMB data. Once CMB lensing is included, the bound on the combination $\Delta N_{\rm eff} \times a_{\rm D}$ is approximately $50\;\mathrm{Mpc}^{-1}$, while $\Delta N_{\rm eff} < 0.33$ (both at 95\% C.L.). The inclusion of DESI BAO tightens the constraint on $\Delta N_{\rm eff} \times a_{\rm D}$ to $1.2\;\mathrm{Mpc}^{-1}$, while the bound on $\Delta N_{\rm eff}$ is slightly relaxed to $0.48$~(95\% C.L.). Indeed, the constraints on $\Delta N_{\rm eff}$ are dominated by the CMB damping tail and are consistent with expectations in the absence of DM-DR interactions ($\Delta N_{\rm eff}< 0.47$ at 95\% C.L. for $a_{\rm D}=0$). Along the same lines, the interaction does not significantly modify the well-known geometric correlations between $\Delta N_{\rm eff}$, $H_0$ and $\omega_{\chi}$, which are still present here. The $\chi^2$ analysis does not show any significant deviations in the goodness of fit of either the CMB or the BAO data due to the recoupling. When the sum of the neutrino masses is also varied, no correlation is observed between $\sum m_\nu$ and $a_{\rm D}$; the upper bound on $\sum m_\nu$ is $\sum m_\nu< 0.079$ eV (95\% C.L.), slightly larger than what is obtained in $\Lambda$CDM~\cite{Elbers:2025vlz}. 

\begin{table}[t]
    \begin{subtable}{\textwidth}
    \centering
    %\resizebox{0.9\textwidth}{!}{
    \begin{tabular}{lcc}
%    \hline 
    \noalign{\vskip 0.5ex}
   % & \multicolumn{3}{c}{$n=-1$}\\[0.5ex]
 & CMB + DESI BAO & CMB(P-ACT) + DESI BAO  \\[0.5ex]
\hline 
\noalign{\vskip 0.5ex}
$\omega_\chi $ & $0.1223 \pm 0.0026$ & $0.1187 \pm 0.0023$   \\[0.5ex]
$\Delta N_{\rm eff}$ & $<0.476$ & $<0.135$ \\[0.5ex]
$\Delta N_{\rm eff} \times (a_{\rm D}/\mathrm{Mpc}^{-1})$ & $<1.16$ & $<12.44$  \\[0.5ex]
$H_0 \, [{\rm km}/{\rm s}/{\rm Mpc}]$ & $69.61 \pm  0.88$ & $68.72 \pm 0.65$ \\[0.5ex]
$\sigma_8$ & $0.8092 \pm 0.0092$ & $0.8014 \pm 0.0200$  \\[0.5ex]
$10^{-4}H_0\hspace{0.2mm} r_{\rm d}$ & $1.0108 \pm 0.0054$ & $1.0097 \pm 0.0088$  \\[0.5ex]
$\Omega_{\rm m}$ & $0.3001 \pm 0.0041$ & $0.3007 \pm 0.0072$  \\[0.5ex]
$S_8$ & $0.8095 \pm 0.0098$ & $0.8028 \pm 0.0164$ \\[0.5ex]
%$\sum m_\nu \, [{\rm eV}]$ & $0.06$ & $0.06$ & $0.06$\\[0.5ex]
\hline 
\noalign{\vskip 3ex}%{\vskip 0.5ex}
    \end{tabular}
    %}
    \end{subtable}
    \begin{subtable}{\textwidth}
    \centering \vspace{-1mm}
    \makebox[\textwidth][c]{%
        \begin{tabular}{lccc}
%        \hline 
        \noalign{\vskip 0.5ex}
        & \multicolumn{3}{c}{CMB + DESI BAO}\\[0.5ex]
        & $a_{\rm D}=0$  &   $\sum m_\nu$ &   $\omega_\chi=0.1\hspace{0.3mm}\omega_{\rm dm}$\\[0.5ex]
\hline 
\noalign{\vskip 0.5ex}
$\omega_{\rm dm}$ & $0.1223  \pm 0.0026$ & $-$ & $0.1224  \pm 0.0026$\\[0.5ex]
$\omega_\chi$ & $-$ & $0.1220 \pm  0.0025$ & $-$\\[0.5ex]
$\Delta N_{\rm eff}$ & $<0.473$ & $<0.444$ & $<0.479$ \\[0.5ex]
$\Delta N_{\rm eff} \times (a_{\rm D}/\mathrm{Mpc}^{-1})$  & $-$ & $<1.38  $ & $<13.08$\\[0.5ex]
$H_0 \, [{\rm km}/{\rm s}/{\rm Mpc}]$ & $69.69 \pm 0.84 $ & $69.55 \pm 0.84$ & $69.63 \pm  0.88$\\[0.5ex]
$\sigma_8$ & $0.8161 \pm 0.0064$ & $0.8126 \pm 0.0100$ & $0.8094 \pm 0.0090$ \\[0.5ex]
$10^{-4}H_0\hspace{0.2mm} r_{\rm d}$ & $1.0116 \pm  0.0052$ & $1.0119 \pm  0.0054 $ &  $1.0108 \pm 0.0052$\\[0.5ex]
$\Omega_{\rm m}$  & $0.2995 \pm 0.0039$ & $0.2992 \pm 0.0041$ & $0.3002 \pm 0.0040$ \\[0.5ex]
$S_8$  & $0.8154\pm0.0076$ & $0.8112\pm0.0101$ & $0.8096 \pm  0.0093$\\[0.5ex]
$\sum m_\nu \, [{\rm eV}]$ & $0.06$ & $<0.079 $ & $0.06$\\[0.5ex]
        \hline 
        \noalign{\vskip 0.5ex}
        \end{tabular}
    }
    \end{subtable}
    \caption{Mean value and marginalized $1\sigma$ error for a selection of the cosmological parameters. When the posterior is one-sided, we quote the 95\%~C.L. upper bound. {\it Top:} The results obtained by leaving out (left) and including (right) the ACT temperature and polarization data. {\it Bottom:} The results obtained by fitting the baseline data combination CMB + DESI BAO to different model variations: a scenario where the fluid DR component does not couple to DM~(left); the DM recoupling model with varying neutrino masses (middle); and a recoupling model where only $10\%$ of DM interacts with DR (right).}
    \label{tab:results}
\end{table}

Including also ACT primary CMB data (orange contours in Fig.~\ref{fig:lin_bounds}) weakens the constraint on the effective coupling, $\Delta N_{\rm eff} \times a_{\rm D} \lesssim 12 \;\mathrm{Mpc}^{-1}$ at 95\% C.L., while at the same time reducing the allowed amount of extra radiation, $\Delta N_{\rm eff} \lesssim 0.14$. Notice in particular the extended tail at large $\Delta N_{\rm eff} \times a_{\rm D}$ but almost constant $\sigma_8$.
This counter-intuitive feature can be understood in the following way. On the one hand, ACT CMB data favor values of $N_{\rm eff} \lesssim 3$ at the 2$\sigma$ level \cite{AtacamaCosmologyTelescope:2025nti}. This implies that requiring $\Delta N_{\rm eff} >0$, consistent with the fact that the DR considered here does not couple to the Standard Model, pushes the posterior of $\Delta N_{\rm eff}$ very close to zero, with a large support at $\Delta N_{\rm eff} \lesssim 10^{-2}$. From Eq.~\eqref{eq:ks} we have $k_{s,\rm{DR}}^2 \propto (\Delta N_{\rm eff}\times a_{\rm D})/\Delta N_{\rm eff}$, such that for very small amounts of DR, and large values of $\Delta N_{\rm eff} \times a_{\rm D}$, the onset of the suppression in the matter power spectrum is pushed to smaller scales, outside of the range of wavenumbers probed by the datasets used in this work. Since the value of $\sigma_8$ is mostly sensitive to $k\lesssim 1 \kMpc$,  the amplitude of clustering remains approximately constant if $\Delta N_{\rm eff} \lesssim 10^{-2}$, unless $a_{\rm D} \gg 10^3 \;\mathrm{Mpc}^{-1}$. This opens up the possibility of testing DM recoupling models with very small DR densities but large interaction strengths using small-scale probes, such as the Lyman-$\alpha$ forest. 
However, it should be noted that for posterior distributions so sharply peaked around the prior boundary, the choices of prior and parametrization themselves, $\Delta N_{\rm eff} \times a_{\rm D}$ vs.~$\log\hspace{0.2mm} (\Delta N_{\rm eff} \times a_{\rm D})$ being one example, could change the final constraints. The apparent cut-off of the orange ellipses in some planes of Fig.~\ref{fig:lin_bounds}, for instance $(\omega_\chi, \Omega_{\rm m})$, is due to the boundary $\Delta N_{\rm eff}>0$.

The credible levels for the model parameters are reported in Table~\ref{tab:results}. As widely appreciated in the literature \cite{Schoneberg:2021qvd}, extra radiation leads to a higher value of the Hubble constant compared to a $\Lambda$CDM fit, but still short of the high value reported by the local distance ladder measurement of SH0ES~\cite{Riess:2025chq}. In this respect, DM recoupling does not significantly alleviate the $H_0$ tension. Finally, we repeat the cosmological analysis assuming only $10\%$ of DM scatters onto DR. The bound on the interaction strength weakens by roughly a factor of 10, see the rightmost column in the bottom part of Table~\ref{tab:results}, consistent with the discussion in Sec.~\ref{sec:linear_theory}.

Our main results can be summarized by saying that in a DM recoupling scenario with $n=-1$, the totality of DM has to be very weakly coupled, $\Gamma_{\chi\text{-}\rm{DR}}/H_0 \lesssim 0.04$ (though note that including ACT primary CMB data relaxes the bound to $\lesssim 0.40$). This constitutes a new test of the collisionless nature of DM on cosmological scales. From a complementary perspective, a $\lesssim 4 \%$ fraction of DM could still be interacting strongly with DR today, i.e.~have~$\Gamma_{\chi\text{-}\mathrm{DR}}/H_0 \gtrsim 1$, with potentially far-reaching consequences for the shape of the matter and galaxy power spectra at low redshift.

\section{Implications for the Microscopic Model}\label{sec:implications_model}
We now discuss the implications of the constraints reported in Sec.~\ref{sec:data} for the microscopic model with Lagrangian in Eq.~\eqref{eq:Lag}. The presentation is divided into two parts. First, we identify the region of parameter space where the discussion in the previous sections applies directly to the dynamics of $\phi$ in the late Universe. Our analysis of the thermal history of the model leads to the requirement of a sufficiently small quartic coupling $\lambda_\phi$, which in turn implies an upper bound on $g_\phi/m_\phi \lesssim \lambda_{\phi}^{1/2}$. Then, focusing on the so-identified region of parameters, we assess the impact of our cosmological constraints by making use of the mapping between phenomenological and microscopic quantities in Eq.~\eqref{eq:macro_micro}. 

\subsection{Identifying the relevant subspace of parameters}\label{sec:subspace_params}
Starting our analysis from the tree-level scalar potential in Eq.~\eqref{eq:Lag}, we take $g_{\phi}>0$ without loss of generality and $\lambda_{\phi}>0$ so that the potential is bounded from below. While a priori $g_\phi$ can take arbitrary values, we restrict our attention to parameters for which the potential has a global minimum at $\langle\phi\rangle = 0$, so that $\phi$ can be directly identified with the physical scalar excitation at zero temperature. Imposing that the global minimum lie at the origin requires
\begin{equation}\label{eq:gphi_max}
g_{\phi}^2 < 3\lambda_{\phi}m_\phi^2\,,
\end{equation}
which in turn enforces $m_\phi^2>0$. The complementary region characterized by $g_{\phi}^2 > 3\lambda_{\phi}m_\phi^2$ is discussed in Appendix~\ref{sec:gphi2_large}, where we show that its features are fully captured (modulo $\mathcal{O}(1)$ numbers) by taking $g_\phi/m_\phi \sim \lambda_\phi^{1/2}$ in Eq.~\eqref{eq:gphi_max}. We require perturbative values for the couplings that appear in the Lagrangian.\footnote{A minimal ultraviolet (UV) origin for Eq.~\eqref{eq:Lag} would be $\mathcal{L} = \mathcal{L}_{\rm kin}(\Phi, \chi) - \mu_\Phi^2 \Phi^2 - \lambda_\Phi \Phi^4/4! - y_\chi \Phi \overline{\chi}\chi$, where a $\mathbb{Z}_2$ symmetry under which $\Phi \to - \Phi$ and $\chi_{L,R} \to \pm \chi_{L,R}$ forbids a mass term for $\chi$ and a $\Phi^3$ interaction. For $\mu_\Phi^2 < 0$, the DM mass and scalar cubic coupling are generated via spontaneous symmetry breaking. However, the predicted value of the cubic coupling is too small to be phenomenologically relevant: expanding $\Phi = \langle \Phi \rangle + \phi$ one finds $g_\phi/m_\phi \sim y_\chi m_\phi/m_\chi$, very suppressed by the ratio $m_\phi/m_\chi$. In this paper we treat all the parameters in Eq.~\eqref{eq:Lag} as independent.}

Next, we must account for thermal corrections, which modify the effective potential for the scalar~\cite{Dolan:1973qd,Weinberg:1974hy}. Since we are interested in the late-Universe dynamics, thermal $\chi$ loops are negligible, being suppressed by $e^{-\, m_\chi(\phi)/T_\chi}$, where the field-dependent masses are
\begin{equation}\label{eq:field_dep_masses}
m_\chi(\phi) \equiv m_\chi + y_\chi \phi\,, \qquad m_\phi^2(\phi) \equiv m_\phi^2 + g_\phi \phi + \frac{\lambda_\phi}{2}\phi^2\,.
\end{equation}
Conversely, the potential originating from $\phi$ loops can be expanded for $T_{\rm DR} \gg m_\phi (\phi)$. After including thermal-loop daisy resummation~\cite{Parwani:1991gq, Arnold:1992rz}, we obtain (see Appendix~\ref{app:Thermal_details} for details)
\begin{equation}\label{eq:V_thermal} \bigg[V_{\rm tree}(\phi) + V^{T}_{1\text{-}\mathrm{loop}} (\phi)\bigg]_{\rm ren} \simeq \frac{1}{2}m_\phi^2 \phi^2 + \frac{g_\phi}{3!}\phi^3 + \frac{\lambda_\phi}{4!}\phi^4 + \frac{T_{\rm DR}^2}{24} m_\phi^2 (\phi) - \frac{T_{\rm DR}}{12\pi} \bigg[m_\phi^2 (\phi) + \frac{\lambda_\phi T_{\rm DR}^2}{24}\bigg]^{3/2}\,, 
\end{equation}
where we have dropped some field-independent pieces. The couplings entering Eq.~\eqref{eq:V_thermal} have been renormalized to remove the UV divergences arising from the zero-temperature $1$-loop potential~\cite{ Coleman:1973jx, Anderson:1991zb}. The critical temperature $T_{\rm c}$, defined as the temperature at which the leading $\sim T_{\rm DR}^2$ correction to the $\phi^2$ term becomes comparable to the zero-temperature contribution, is therefore
\begin{equation}
T_{\rm c}^2 \equiv 24\,\frac{m_\phi^2}{\lambda_{\phi}}\,.
\end{equation}
At temperatures $T_{\rm DR} \gg T_{\rm c}$, thermal corrections dominate, shifting the position of the global minimum away from the origin and altering the mass and couplings of the physical excitation relative to the zero-temperature potential. Expanding the theory around the temperature-dependent minimum, $\phi = \langle \phi \rangle_{T} + \widehat{\phi}$ where
\begin{equation}
     \langle\phi\rangle_{T} \simeq -\,\frac{g_{\phi}
     }{\lambda_{\phi}} \Bigg[ 1 - \frac{8 \big(3 \lambda_{\phi} m_\phi^2 - g_{\phi}^2\big)}{ \lambda_{\phi}^2 T_{\rm DR}^2}  + \mathcal{O}\bigg(\frac{T^4_{\rm c}}{T_{\rm DR}^4} \bigg) \Bigg]\,,
\end{equation}
one finds that a thermal mass $\sim T_{\rm DR}^2$ is generated, while the cubic coupling becomes suppressed as $\sim 1/T_{\rm DR}^2\,$,
\begin{align}\label{eq:mass_cubic_highT}
    m_{\widehat{\phi}, T}^2\simeq&\;  \frac{\lambda_\phi T_{\rm DR}^2}{24} + m_\phi^2 - \frac{ g_\phi^2}{2\lambda_\phi} \Bigg[ 1 + \mathcal{O}\bigg( \frac{T_{\rm c}^4}{T_{\rm DR}^4}\bigg) \Bigg]\,, \\
    g_{\widehat{\phi}, T}  \simeq&\; g_{\phi}\Bigg[ \frac{8\,   \big(3 \lambda_{\phi} m_\phi^2 - g_{\phi}^2\big)}{
 \lambda_{\phi}^2 T_{\rm DR}^2} + \mathcal{O}\bigg(\frac{T^4_{\rm c}}{T_{\rm DR}^4} \bigg) \Bigg]\,.
\end{align}
The quartic coupling is just $\lambda_{\widehat{\phi},T} \simeq \lambda_\phi$, up to negligible corrections suppressed by powers of $\lambda_\phi^{1/2}\big/\pi \ll 1$. As a result, at high temperatures $T_{\rm DR} \gg T_{\rm c}$ thermal corrections strongly modify the redshift dependence of the momentum-transfer rate in Eq.~\eqref{eq:Gamma_g2y2}, yielding the very steep scaling $n=-\,6$ in Eq.~\eqref{eq:Gamma_DR_chi_general}. 

The above discussion shows that the cosmological analysis performed in the previous sections, which assumed $n = -\,1$, applies to the region of parameter space where the critical temperature is high enough that the DM recoupling dynamics takes place at $T_{\rm DR} \ll T_{\rm c}$. In this regime, the scalar potential is well approximated by its zero-temperature form. We thus require that $T_{\rm c} > T_{{\rm DR},\hspace{0.2mm} z_\ast}$ for a sufficiently high $z_\ast$. Taking $1+z_\ast = 100$ as reference, we obtain
\begin{equation}\label{eq:lambdaphi_upperbound}
    \lambda_\phi < 1.7 \times 10^{-5}\,  \bigg( \frac{m_\phi }{10^{-5}\;\mathrm{eV}} \bigg)^2 \left( \frac{0.5 }{\xi_0} \right)^2  \left( \frac{100 }{1 + z_*} \right)^2 \,.
\end{equation}
In conclusion, taking a sufficiently small $\lambda_\phi$ ensures that the constraints derived in Sec.~\ref{sec:data} apply directly to the parameters of the Lagrangian in Eq.~\eqref{eq:Lag}. Importantly, this is safely compatible with the requirement that DR remains strongly coupled with itself at all times, see Appendix~\ref{app:DR_selfscatt}. A maximum size for $\lambda_\phi$ implies, in turn, a maximum size for $g_\phi$, given the condition in Eq.~\eqref{eq:gphi_max} to have a global minimum at the origin at zero temperature.

\subsection{Setting constraints}
\begin{figure}
    \centering
\includegraphics[width=0.75\linewidth]{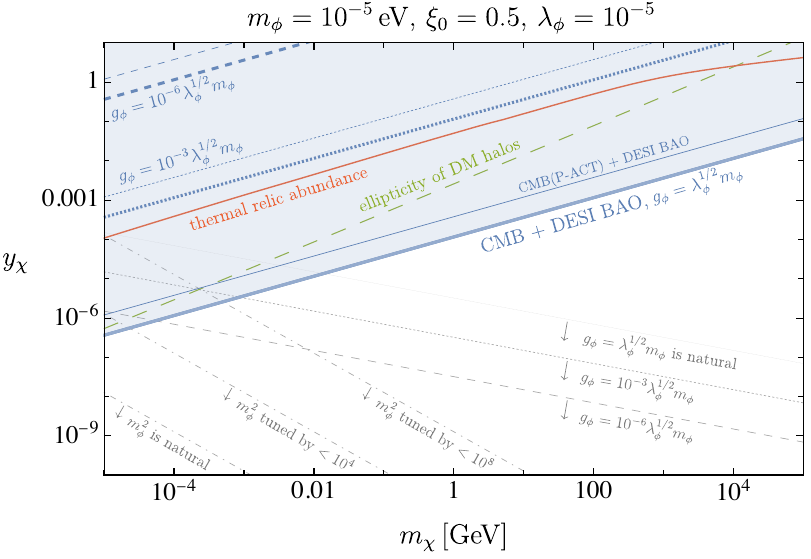}
    \caption{Impact of our cosmological constraints on the parameter space of the microscopic dark sector model with Lagrangian in Eq.~\eqref{eq:Lag}. We have fixed the DR parameters to $m_\phi = 10^{-5}\;\mathrm{eV}$ and $\xi_0 = 0.5$. The quartic coupling has been set to $\lambda_\phi = 10^{-5}$, so that the critical temperature $T_{\rm c} \equiv m_\phi / (\lambda_\phi/24)^{1/2}$ is high enough that the recoupling dynamics takes place at $T_{\rm DR} \ll T_{\rm c}$. The region above the {\color{lightblue}\bf thick~(solid, dotted, dashed)~blue lines} is excluded at $95\%$~C.L. by our combination of CMB + DESI BAO data, assuming $g_\phi = (1, 10^{-3}, 10^{-6})\lambda_\phi^{1/2} m_\phi$, while {\color{lightblue}\bf thin blue lines} correspond to CMB(P-ACT) + DESI BAO, for the same values of $g_\phi$. On the {\color{lightred}\bf red curve}, $\chi$ obtains the observed DM abundance via thermal freeze out of $\chi\overline{\chi}\to \phi\phi$ annihilation. The {\bf\color{lightgreen} long-dashed green line} indicates the astrophysical constraint on DM self-interactions, estimated from the observed ellipticity of galactic DM halos~\cite{Feng:2009mn,Agrawal:2016quu,McDaniel:2021kpq}. {\color{lightgray}\bf Dot-dashed gray lines} are fine-tuning contours for $m_\phi^2$. {\color{lightgray}\bf (Solid, dotted, dashed) gray lines} are fine-tuning contours for $g_\phi$, which is assumed to take the value corresponding to the pair of blue lines with matching pattern. Only fine-tuning related to DM loops is considered, see the text for more details.}
    \label{fig:micro_param_space}
\end{figure}

An exemplary parameter subspace that satisfies the conditions laid out in Sec.~\ref{sec:subspace_params} is shown in Fig.~\ref{fig:micro_param_space}. We have chosen $m_\phi = 10^{-5}\;\mathrm{eV}$ and $\xi_0 = 0.5$ (corresponding to $\Delta N_{\rm eff} \approx 0.14$) as representative DR parameters, and consequently we have set $\lambda_\phi = 10^{-5}$, just below the largest value allowed by Eq.~\eqref{eq:lambdaphi_upperbound}. We are left with three free parameters, namely $y_\chi$, $m_\chi$ and $g_\phi$. In Fig.~\ref{fig:micro_param_space} we show $y_\chi$ vs.~$m_\chi$, while $g_\phi$ is fixed to representative values. We discuss here the salient findings, postponing many details to Appendix~\ref{app:Thermal_details}.

The red curve shows the value of $y_\chi$ for which thermal freeze out of $\chi\overline{\chi} \to \phi\phi$ annihilation in the early Universe~\cite{An:2016kie} yields the observed DM abundance, see Appendices~\ref{app:DM_ann} and~\ref{app:DM_relic_density}. The curve follows $y_\chi \propto m_\chi^{1/2}$ except at the largest DM masses, where Sommerfeld enhancement becomes important~\cite{Cassel:2009wt}. The thermal abundance curve is essentially independent of $g_\phi$. For the DM mass, we consider the range $10\;\mathrm{keV} < m_\chi < 100\;\mathrm{TeV}$ (corresponding to $10^{-9} > m_\phi/m_\chi > 10^{-19}$). The upper end is the approximate unitarity limit for thermal freeze out, where $y_\chi$ approaches $4\pi$. The lower end cuts out the $m_\chi \sim \mathcal{O}(\mathrm{keV})$ region, where the DM Jeans wavenumber $k_{s,\chi}$ (see the right panel of Fig.~\ref{fig:TDM}) is small enough to give a suppression of power on scales probed by complementary astrophysical observables, such as the Lyman-$\alpha$ flux, strong galaxy lensing and Milky Way satellite counts. 

The thick blue lines in Fig.~\ref{fig:micro_param_space} project on the $(m_\chi, y_\chi)$ plane the constraint we derived from the baseline CMB + DESI BAO analysis, $\Delta N_{\rm eff} \times (a_{\rm D}/\mathrm{Mpc}^{-1}) < 1.16$ (top left column in Table~\ref{tab:results}), for three different assumptions on the size of the scalar cubic self-coupling: $g_\phi/m_\phi = \lambda_{\phi}^{1/2}$, which lies very close to the maximum value allowed by Eq.~\eqref{eq:gphi_max}, as well as coupling strengths smaller by factors $10^3$ and $10^6$. Remarkably, we find that, for the chosen representative values of $m_\phi$ and $\xi_0$, CMB + DESI BAO data rule out the thermal freeze-out curve for cubic self-coupling strengths
\begin{equation}
 g_\phi/m_\phi \gtrsim 10^{-5} \,.
\end{equation}
The sensitivity to the thermal DM parameter space could be extended to even smaller $g_\phi$, by exploiting upcoming galaxy survey data. The thin blue lines show the weaker constraints resulting from our analysis of the CMB(P-ACT) + DESI BAO dataset, $\Delta N_{\rm eff} \times (a_{\rm D}/\mathrm{Mpc}^{-1}) < 12.44$ (top-right column in Table~\ref{tab:results}).

We note that DM is still relativistic at the epoch of Big Bang Nucleosynthesis (BBN) if $m_\chi \ll \xi_{\rm BBN} T_{\gamma,\mathrm{BBN}}$, where $T_{\gamma,\mathrm{BBN}} \sim 1\; \mathrm{MeV}$. Nevertheless, even in this light DM region the predicted $\Delta N_{\rm eff}$ at BBN is smaller than the one at CMB decoupling. Conservation of entropy yields~\cite{Ackerman:2008kmp}
\begin{equation}
\Delta N_{\rm eff, BBN} = \frac{4}{7}\hspace{0.4mm} g_{\ast \mathrm{DS,BBN}} \left( \frac{g_{\ast s\, \mathrm{vis,BBN}}}{g_{\ast s\, \mathrm{vis,0}}}\, \frac{g_{\ast s\, \mathrm{DS,0}}}{g_{\ast s\, \mathrm{DS,BBN}}} \right)^{4/3} \xi_0^4 \,,
\end{equation}
where $g_{\ast s\, \mathrm{vis,BBN}} = 10.75$ and $g_{\ast s\, \mathrm{vis,0}} = 3.91$. For $m_\chi \ll \xi_{\rm BBN}\hspace{0.3mm} \mathrm{MeV} \approx 0.85\,\xi_0\, \mathrm{MeV}$, one has $g_{\ast \mathrm{DS,BBN}} = g_{\ast s\, \mathrm{DS,BBN}} = 4.5$ and $g_{\ast s\, \mathrm{DS,0}} = 1$, resulting in $\Delta N_{\rm eff, BBN} \approx 1.3\,\xi_0^4 \approx 0.61\, \Delta N_{\rm eff}$, where $\Delta N_{\rm eff}$ is the value at CMB decoupling in Eq.~\eqref{eq:oDR_Neff}. Hence, satisfying the late-Universe constraints we derive in this work automatically guarantees consistency with BBN (see e.g.~Ref.~\cite{Schoneberg:2024ifp} for a recent review). For instance, $\xi_0 = 0.5$ as chosen in Fig.~\ref{fig:micro_param_space} gives $\Delta N_{\rm eff, BBN} \approx 0.083$ in the sub-MeV region for the DM mass.  

For the sake of comparison, in Fig.~\ref{fig:micro_param_space} we also report (long-dashed green line) the strongest astrophysical constraint on the long-range DM self-interaction mediated by Yukawa exchange of the light mediator $\phi$, derived from the observed ellipticity of the DM halo of the NGC720 galaxy~\cite{Feng:2009mn,Agrawal:2016quu}. This constraint is independent of $g_\phi$. Although the ellipticity constraint was later extended to a sample of $11$ galaxies~\cite{McDaniel:2021kpq}, it is subject to important caveats~\cite{Agrawal:2016quu,McDaniel:2021kpq} and should rather be regarded as an illustration of the potential complementarity between cosmological precision measurements and small-scale observables in probing the nature of DM.

Finally, we consider the naturalness of our parameter space. Since the DM-DR coupling $y_\chi$ generates radiative contributions to the scalar potential, naturalness requires
\begin{equation}
m_\phi^2 \gtrsim \frac{y_\chi^2}{(4\pi)^2}\,m_\chi^2\,, \qquad g_\phi \gtrsim \frac{y_\chi^3}{(4\pi)^2}\,m_\chi\,,\qquad \lambda_\phi \gtrsim \frac{y_\chi^4}{(4\pi)^2}\,,
\end{equation}
where the loop momenta were cut off at $m_\chi$. Interestingly, we find that cosmological data have already probed a wedge of parameter space where $g_\phi$ has a natural size, extending up to $m_\chi \sim 10\;\mathrm{MeV}$, as shown in Fig.~\ref{fig:micro_param_space} by the overlap of the solid blue and solid gray lines corresponding to $g_\phi/m_\phi = \lambda_\phi^{1/2}$. In this wedge, however, $m_\phi^2$ requires a fine-tuning of at least $\sim 10^3$, as shown by the dot-dashed gray lines. Naturalness of the quartic coupling only requires $y_\chi \lesssim 0.2$ (not shown) for our chosen value $\lambda_\phi = 10^{-5}$, which is satisfied in most of the plane.   

Additional fine-tuning of the scalar mass can arise from loops of $\phi$ itself. The contribution of the cubic self-coupling is natural provided $m_\phi^2 \gtrsim g_\phi^2 / (4\pi)^2$, which is always satisfied since $g_\phi^2 \lesssim \lambda_\phi m_\phi^2$ and $\lambda_\phi$ is perturbative. However, the contribution of the quartic coupling requires $m_\phi^2 \gtrsim \lambda_\phi\hspace{0.2mm} \Lambda_{\rm UV}^2 / (4\pi)^2$, where $\Lambda_{\rm UV}$ is the UV cutoff. Assuming $\Lambda_{\rm UV} = m_\chi$ would lead to $m_\phi^2$ being fine-tuned by at least $10^{10}$ in all the plane shown in Fig.~\ref{fig:micro_param_space}, but smaller values of the cutoff are possible depending on how the model is completed in the UV. Since quantifying this contribution to the fine-tuning requires additional assumptions, we do not show it in Fig.~\ref{fig:micro_param_space}, focusing only on the fine-tuning arising directly from DM loops.

\section{Outlook}\label{sec:outlook}
In this work we investigated, for the first time, the possibility that DM interacts only at low redshift with a relativistic dark radiation species. We dubbed this dynamics dark matter recoupling, as the ratio between the DM momentum transfer rate and the Hubble parameter grows with time. We studied in detail a recoupling scenario where the DM momentum transfer rate is constant in time ($n= - 1\hspace{0.2mm}$ in the phenomenological parametrization introduced in Sec.~\ref{sec:intro}), leading to a distinctive and fast-growing scale-dependent suppression of the matter power spectrum, starting typically after reionization. Such late-time phenomenology implies that the primary CMB power spectrum is mostly unaffected by the new interaction, and DM recoupling can only be tested with Large Scale Structure (LSS) observables like lensing and galaxy clustering, as discussed in Sec.~\ref{sec:cosmo}. 

DM recoupling is easily realized in microscopic dark sector scenarios, as we showed in Sections~\ref{sec:DM_recoupling} and \ref{sec:implications_model}. Here we considered a simple model where fermionic DM couples to scalar DR. The DM-DR Yukawa interaction involved in the recoupling dynamics can also mediate DM production via thermal freeze out during radiation domination, leading to a non-trivial interplay between early Universe physics and late-time cosmological probes. We found that, while the scalar cubic self-coupling that is essential for the recoupling mechanism can be natural in the observationally accessible parameter space, the scalar DR mass requires fine-tuning.

The constraints we presented in Sec.~\ref{sec:data} indicate that all of DM cannot be fully recoupled today. Our results, together with the vast body of earlier work on DM decoupling scenarios, establish the approximately collisionless nature of DM on cosmological scales. Nevertheless, current data still allow roughly $4\%$ of DM to be strongly coupled to DR at low redshift. To further improve upon this bound using recent and upcoming high-precision data from DESI and Euclid, the development of new perturbative solutions will be required, going beyond the linear regime discussed here. We will return to this important task in future work. Furthermore, our analysis was limited to the simplest setup where DM recoupling is realized, leaving open the broader investigation of this class of models.

\section*{Acknowledgments}
E.D., E.C. and E.S. thank the Galileo Galilei Institute for Theoretical Physics for hospitality and the INFN for partial support during the completion of this work. E.D. and E.S. were supported by the grant RYC2023-042775-I, funded by the Spanish Ministry of Science, Innovation and Universities (MCIU) through the Spanish State Research Agency (AEI, 10.13039/501100011033) and by the FSE+. The work of E.D. was carried out within the framework of the Doctoral Program in Physics at the Universitat Aut\`onoma de Barcelona.

\appendix

\section{Dark Matter-Dark Radiation Momentum Transfer Rate}\label{app:mom_transf_rate}
In this appendix we summarize our calculation of the momentum transfer rate for the scattering between DM and DR. Our definition of the (conformal) DM momentum transfer rate~\cite{Cyr-Racine:2015ihg} is $\Gamma_{\chi\text{-}\mathrm{DR}} =  a\hspace{0.2mm} \gamma_{\chi\text{-}\mathrm{DR}}/2$, where $\gamma_{\chi\text{-}\mathrm{DR}}$ is given by (see e.g.~Ref.~\cite{Bringmann:2016ilk})
\begin{equation}
\gamma_{\chi\text{-}\mathrm{DR}} = \frac{1}{384\pi^3 \eta_\chi T_{\rm DR} m_\chi^3} \int \hspace{-1mm}\mathrm{d}E \big[1 \mp f_{\rm DR}^{\pm} (E)\big]f_{\rm DR}^{\pm}(E) \hspace{-1mm}\int_{-4 p^2}^0 \mathrm{d}t (-t) \sum_{\rm spins}|\mathcal{M}|^2 (s, t)\,.
\end{equation}
For generality, in this expression we allow DR to be either bosonic or fermionic, $f_{\rm DR}^{\pm}(E) \equiv \big(e^{E/T_{\rm DR}} \pm 1\big)^{-1}$. We now change integration variable from $t$ to $\tilde{\mu}$ in the inner integral, use the relation $\big[1 \mp f_{\rm DR}^{\pm}(E)\big]f_{\rm DR}^{\pm}(E) = -\, T_{\rm DR} \partial_{E} f_{\rm DR}^{\pm}(E)$, and finally we switch variable from $E$ to $p$. We thus arrive at
\begin{equation}
\Gamma_{\chi\text{-}\mathrm{DR}} = -\, \frac{a\hspace{0.2mm} \eta_{\rm DR}}{96\pi^3 m_\chi^3} \int_0^\infty \mathrm{d}p\, \frac{\partial f_{\rm DR}^{\pm}(p)}{\partial p}\, p^4 \big[A_0(p) - A_1(p)\big]\,,
\end{equation}
where the $A_\ell (p)$ projections were defined in Eq.~\eqref{eq: projection onto the lth Legendre polynomial}. Using $\Gamma_{\mathrm{DR}\text{-}\chi} = 3\bar{\rho}_\chi \Gamma_{\chi\text{-}\mathrm{DR}} / (4\bar{\rho}_{\rm DR})$ and selecting bosonic DR, $f_{\rm DR}^{(0)}(p) = f_{\rm DR}^{-}(p)$, we obtain Eq.~\eqref{eq:Gamma DR-DM from ETHOS}.

The calculation of $\sum_{\rm spins}|\mathcal{M}|^2$ for the $\chi (p_1) \phi (p_2) \rightarrow \chi(p_3) \phi(p_4)$ process is straightforwardly performed starting from the matrix element in Eq.~\eqref{eq:scattering amplitude} and recalling the definitions of the Mandelstam variables
\begin{equation}\label{eq:Mandelstam}
    s = -\,(p_1+p_2)^2\,,\qquad t = -\,(p_1-p_3)^2\,, \qquad
    u = -\,(p_1-p_4)^2\,,
\end{equation}
which obey $s+t+u = 2m_{\chi}^2+2m_{\phi}^2$. According to Eq.~\eqref{eq: squared scattering amplitude}, we split the computation into the three pieces arising from the $t$-channel diagram alone, the $s/u$-channel diagrams alone, and their interference. For the $t$-channel $\mathcal{O}(g_\phi^2 y_\chi^2)$ contribution, after integrating over angles we obtain
\begin{align}
\big[ A_0(p) - A_1(p) \big]_{g_\phi^2 y_\chi^2} =&\; - \frac{g_\phi^2 y_\chi^2 m_\chi^2}{2 p^4}  \left[ \frac{4 p^2}{m_\phi^2 + 4 p^2} + \log \frac{m_\phi^2}{m_\phi^2 + 4 p^2}\right] \nonumber \\ &\quad\qquad+ \frac{g_\phi^2 y_\chi^2 }{4 p^4}  \left[\frac{4 p^2 \big( m_\phi^2 + 2 p^2\big)}{m_\phi^2 + 4 p^2} + m_\phi^2 \log \frac{m_\phi^2}{m_\phi^2+4 p^2}\right].\label{eq: g^2y^2 projection}
\end{align}
Plugging this expression into Eq.~\eqref{eq:Gamma DR-DM from ETHOS}, integrating by parts and setting $m_\phi = 0$ in the $\mathcal{O}\left(1/m_\chi^{3}\right)$ term gives Eq.~\eqref{eq:Gamma_DRchi_intermediate}. For the $s/u$-channel $\mathcal{O}(y_\chi^4)$ contribution we find, expanding for large $m_\chi$,
\begin{align}
\big[ A_0(p) - A_1(p) \big]_{y_\chi^4} =&\; \frac{4 y_\chi^4}{15} \Bigg[ \frac{ 5 p^4 }{ \big(m_\phi^2 + p^2\big)^{2}} - \frac{4 p^6}{m_\chi \big(m_\phi^2 + p^2\big)^{5/2}} \Bigg] + \mathcal{O}\left(1/m_\chi^2\right)\,. \label{eq: y^4 projection}
\end{align}
After substituting into Eq.~\eqref{eq:Gamma DR-DM from ETHOS}, we integrate by parts and take $m_\phi = 0$ in the $\mathcal{O}\left(1/m_\chi^{4}\right)$ piece, arriving at 
\begin{equation}
\Gamma_{\mathrm{DR}\text{-}\chi\,(y_\chi^4)} = \frac{\omega_\chi}{\omega_{\rm DR}}\frac{(1+z)^2}{m_\chi^3}\frac{\pi y_\chi^4}{360}\hspace{0.3mm} T_{\rm DR,0}^4 \left( 1 - \frac{5}{\pi^2}\frac{m_\phi^2}{T_{\rm DR}^2} - \frac{360\zeta(5)}{\pi^4} \frac{T_{\rm DR}}{m_\chi} + \dots\right)\,,
\end{equation}
which is Eq.~\eqref{eq:Gamma_DR_y4}. Finally, the $\mathcal{O}(g_\phi y_\chi^3)$ interference contribution expanded for large $m_\chi$ is
\begin{align}
\big[ A_0(p) -& A_1(p) \big]_{g_\phi y_\chi^3} = -\, \frac{g_\phi y_\chi^3 m_\chi m_\phi^2}{2p^4}\,\left[ \frac{4 p^2}{m_\phi^2 + p^2} + \frac{m_\phi^2 + 2 p^2}{m_\phi^2 + p^2} \log \frac{m_\phi^2}{m_\phi^2 + 4 p^2 } \right] \nonumber \\
-&\; \frac{g_\phi y_\chi^3}{12 p^4}\, \frac{ 4p^2 \big(8 p^4 + 6 m_\phi^2 p^2 + 3 m_\phi^4  \big) + 3 m_\phi^4 \big( m_\phi^2 + 4 p^2 \big) \log \frac{m_\phi^2}{m_\phi^2 + 4 p^2} }{ \big(m_\phi^2+p^2\big)^{3/2}} + \mathcal{O}\left(1/m_\chi\right)\,.
\end{align}
Plugging this into Eq.~\eqref{eq:Gamma DR-DM from ETHOS} and integrating by parts, we obtain
\begin{align}
\Gamma_{\mathrm{DR}\text{-}\chi\,(g_\phi y_\chi^3)} \simeq -\, \frac{\omega_\chi}{\omega_{\rm DR}} \frac{(1+z)^{-2}}{128\pi^3 m_\chi^2}g_\phi y_\chi^3 \Bigg[  m_\phi^2\hspace{-1mm} \int_0^\infty \hspace{-1.3mm} \mathrm{d}p\hspace{0.3mm} p &f_{\rm DR}^{(0)}(p)\, \frac{4 p^2 \big(m_\phi^2 - 2 p^2\big) + m_\phi^2 \big(m_\phi^2 + 4 p^2\big) \log \frac{m_\phi^2}{m_\phi^2 + 4 p^2} }{\big(m_\phi^2+p^2\big)^2\, \big(m_\phi^2 + 4 p^2\big)} \nonumber \\
+&\, 16\hspace{0.2mm} \zeta(3) \frac{T_{\rm DR}^3}{m_\chi} \Bigg]\,, \label{eq:gy3_intermediate}
\end{align}
where in the $\mathcal{O}\big(1/m_\chi^3\big)$ term we have set $m_\phi = 0$. Our work is not finished yet, though: the integrals in Eqs.~\eqref{eq:Gamma_DRchi_intermediate} and~\eqref{eq:gy3_intermediate} diverge for $m_\phi \to 0$, hence special care must be taken in evaluating them. We turn to that task next.

\subsection{Evaluating integrals that diverge in the limit of massless dark radiation}\label{sec:regions}
The calculation of $\Gamma_{\mathrm{DR}\text{-}\chi\,(g_\phi^2 y_\chi^2)}$ requires the evaluation of the integral in Eq.~\eqref{eq:Gamma_DRchi_intermediate}, which diverges for $m_\phi \to 0$. To obtain its asymptotic behavior at small $m_\phi/T_{\rm DR}$, we first change integration variable to $x \equiv E_{\rm DR}/T_{\rm DR}$,
\begin{equation}
I(z) \equiv  \int_z^\infty  \frac{\mathrm{d}x}{e^x - 1}\, \frac{x (x^2 - z^2)}{\big( x^2 - \frac{3}{4} z^2\big)^2}\,, \qquad z \equiv \frac{m_\phi}{T_{\rm DR}}\,.
\end{equation}
Following a procedure proposed in Ref.~\cite{Rubira:2022xhb} by adapting the method of regions~\cite{Beneke:1997zp}, we split the integral in two as
\begin{equation}
I(z) = \int_{z}^\lambda (\ldots) + \int_{\lambda}^\infty (\ldots) = I_1^\lambda(z) + I_2^{\lambda}(z)\,,
\end{equation}
where we have introduced an arbitrary cutoff parameter $\lambda$, satisfying $z \ll \lambda \ll 1$. The idea is that in $I_1^\lambda(z)$ we can expand the Bose-Einstein distribution for $x\ll 1$,
\begin{equation}
\frac{1}{e^x - 1} = \frac{1}{x}-\frac{1}{2}+\frac{x}{12}-\frac{x^3}{720}+\frac{x^5}{30240}+\mathcal{O}\left(x^7\right),
\end{equation}
and perform analytically the resulting integrals of ratios of polynomials. We then expand for small $z$, obtaining
\begin{equation}\label{eq:I1_g2y2}
I_1^\lambda(z) =   \frac{\mathcal{C}_{g^2_\phi y_\chi^2}}{z} +\frac{\log z}{2} +\frac{1}{4} -\frac{1}{2} \log 2 + f(\lambda) + \mathcal{O}(z)\,,
\end{equation}
where $f(\lambda)\equiv -\frac{1}{\lambda} -\frac{1}{2} \log \lambda +\frac{\lambda}{12} -\frac{\lambda^3}{2160} + \frac{\lambda^5} {151200} + \mathcal{O}(\lambda^7)$ and $\mathcal{C}_{g_\phi^2 y_\chi^2} \equiv \big[7 \sqrt{3} \log \left(2+ \sqrt{3}\right) - 6\big] \big/ 9$. On the other hand, in $I_2^\lambda(z)$ we are allowed to expand the integrand for $z \ll x$,
\begin{equation}\label{eq:I2_g2y2}
I_2^\lambda(z) = \int_\lambda^\infty  \frac{\mathrm{d}x}{e^x - 1}\, \frac{1}{x} + \mathcal{O}(z^2) =  - f(\lambda) - \frac{\log (2\pi) - \gamma_{\rm E}}{2}  + \mathcal{O}(z^2)\,, 
\end{equation}
where $\gamma_{\rm E}$ is the Euler-Mascheroni constant. Summing Eqs.~\eqref{eq:I1_g2y2} and~\eqref{eq:I2_g2y2} the dependence on the arbitrary cutoff $\lambda$ cancels, as it should, and we arrive at
\begin{equation}
I(z) =  \frac{\mathcal{C}_{g^2_\phi y_\chi^2}}{z} +\frac{\log z}{2} +\frac{1}{4} - \frac{\log (4\pi) - \gamma_{\rm E}}{2}  + \mathcal{O}(z)\,.
\end{equation}
The leading $\mathcal{O}(1/z)$ term of this expression was given in Eq.~\eqref{eq:Gamma_g2y2}.

In addition, to obtain the expression of $\Gamma_{\mathrm{DR}\text{-}\chi\,(g_\phi y_\chi^3)}$ we need to evaluate the integral in Eq.~\eqref{eq:gy3_intermediate}, which is also divergent for $m_\phi \to 0$. Changing integration variable to $x = E_{\rm DR}/T_{\rm DR}$, we cast this integral in the form
\begin{equation}
J(z)\equiv \int_z^\infty \frac{\mathrm{d}x}{e^x - 1}\, \frac{  -\, 2 x^4 + 5 x^2 z^2 - 3 z^4 + z^2 \left( x^2 - \tfrac{3}{4} z^2\right) \log \frac{z^2}{4 x^2-3 z^2} }{x^3  \left(x^2 - \tfrac{3}{4} z^2\right)}\;.
\end{equation}
As we did above for $I(z)$, we split $J(z) = J_1^\lambda(z)+J_2^\lambda(z)$ with $z \ll \lambda \ll 1$ and calculate separately the two pieces, expanding the Bose-Einstein distribution for $x \ll 1$ in the former and the integrand for $z\ll x$ in the latter. This yields
\begin{align}
J_{1}^\lambda (z) =&\,   -   \frac{4\hspace{0.3mm}\mathcal{C}_{g^2_\phi y_\chi^2}}{3z} - \log z - 1 + \log 2 - 2 f(\lambda)  + \mathcal{O}(z)\,, \\
J_{2}^\lambda (z) =&\; 2 \left[ f(\lambda) + \frac{\log (2\pi) - \gamma_{\rm E}}{2} \right] + \mathcal{O}(z^2)\,,
\end{align}
and finally
\begin{equation}
J(z) =   -   \frac{4\hspace{0.3mm}\mathcal{C}_{g^2_\phi y_\chi^2}}{3z} - \log z - 1 +   \log (4\pi) - \gamma_{\rm E}   + \mathcal{O}(z)\,,
\end{equation}
where the leading $\mathcal{O}(1/z)$ term was given in Eq.~\eqref{eq:Gamma_gy3}.

\section{Self-Scattering of Dark Radiation}\label{app:DR_selfscatt}
The thermally-averaged, conformal rate for $\phi(p_1)\phi(p_2)\to \phi(p_3)\phi(p_4)$ scattering, $\langle \Gamma_{\phi\phi} \rangle$, is given by~\cite{Oldengott:2014qra}
\begin{align}
n_{\rm DR}^{(0)} \langle \Gamma_{\phi\phi} \rangle = a  \int \prod_{i\,=\,1}^4 \frac{\mathrm{d}^3 p_i}{(2\pi)^3 2E_i} &\, f_{\rm DR}^{\pm}(E_1) f_{\rm DR}^{\pm}(E_2) \big[1 \mp f_{\rm DR}^{\pm}(E_3)\big] \big[1\mp f_{\rm DR}^{\pm}(E_4)\big] \nonumber \\  \times&\, \frac{\sum_{\rm spins} |\mathcal{M}|^2}{N_f !}(2\pi)^4 \delta^D(p_1 + p_2 - p_3 - p_4)\,,  \label{eq:integral_definition_DRscatt}
\end{align}
where initial~(final) momenta are taken ingoing~(outgoing), $N_f = 2$ accounts for the identical particles in the final state, and $n_{\rm DR}^{(0)} = \eta_{\rm DR} \int \frac{\mathrm{d}^3 p}{(2\pi)^3} f_{\rm DR}^{\pm}(E)\,$ is the unperturbed number density of DR. The matrix element is obtained by summing over $4$ Feynman diagrams,
\begin{equation}
\mathcal{M} = -\, g_\phi^2 \Big( \frac{1}{s-m_\phi^2} + \frac{1}{t-m_\phi^2} + \frac{1}{u-m_\phi^2} \Big) - \lambda_\phi\,,
\end{equation}
where $s, t$ and $u$ are defined as in Eq.~\eqref{eq:Mandelstam}, but here they satisfy $s + t + u = 4m_\phi^2$. Since our DR is a scalar, we select the Bose-Einstein distribution in the above equations, $f^{-}_{\rm DR} (E) = f_{\rm DR}^{(0)}(E)$.

In previous literature~\cite{Brinckmann:2022ajr,Chang:2025uvx} the integral on the right-hand side of Eq.~\eqref{eq:integral_definition_DRscatt} was evaluated only for $g_\phi = m_\phi = 0$. Furthermore, several simplifying assumptions were made: the $1\mp f_{\rm DR}^{\pm}$ statistical factors were set to $1$ and the DR distribution function was assumed to take the Maxwell-Boltzmann form. These assumptions were made to simplify the computation, but they are not justified a priori and we do not adopt them here. As we will show, our more accurate treatment significantly alters the result for $\langle \Gamma_{\phi\phi} \rangle$, even in the previously-considered case $g_\phi = m_\phi = 0$.

We parametrize the three-momenta as
\begin{align}\label{eq:3momenta}
    &\vec{p}_1 = p_1 (0, 0, 1) \,,\qquad \vec{p}_2 = p_2 (0, \sin{\alpha},\cos{\alpha}) \,,\qquad \vec{p}_3 = p_3(\sin{\beta} \sin{\theta}, \cos{\beta} \sin{\theta}, \cos{\theta})\,, 
\end{align}
and adapt previous results~\cite{1976Ap&SS..39..429Y,Hannestad:1995rs} to reduce the $12$-dimensional integral on the right-hand side of Eq.~\eqref{eq:integral_definition_DRscatt} to a $5$-dimensional integral,
\begin{align}
&\frac{a}{16\pi^6}  \int_0^\infty \hspace{-1mm} \frac{\mathrm{d}p_1\hspace{0.2mm} p_1^2}{ 2E_1} \int_0^\infty \hspace{-1mm}\frac{\mathrm{d}p_2 \hspace{0.2mm} p_2^2}{2 E_2} \int_0^{p_{3}^{\rm max}} \hspace{-1mm} \frac{\mathrm{d} p_3 \hspace{0.2mm} p_3^2}{2 E_3}\, f_{\rm DR}^{(0)}(E_1) f_{\rm DR}^{(0)}(E_2) \big[1 + f_{\rm DR}^{(0)}(E_3)\big] \big[1 + f_{\rm DR}^{(0)}(E_1+E_2 - E_3)\big] \nonumber\\ 
 \label{eq:general_Gamma}
&\times\hspace{-1mm} \int_{\hat{\alpha}}^{\hat{\beta}} \mathrm{d}\cos \theta \int_{x_-}^{x_+} \mathrm{d}\cos\alpha\, \frac{\sum_{\rm spins} |\mathcal{M}|^2}{N_f !}\bigg|_{\cos\beta_+}\hspace{-1mm} (p_1, p_2, p_3, \cos\theta, \cos\alpha)\, \frac{1}{\sqrt{\hat{a} \cos^2 \alpha + \hat{b} \cos\alpha + \hat{c}}}\,.
\end{align}
On the first line, $p_3^{\rm max} = \sqrt{(E_1 + E_2 - m_\phi)^2 - m_\phi^2}\,$. On the second line, the boundaries of the $\cos\theta$ integration are, defining $\gamma \equiv E_1 E_2 - E_1 E_3 - E_2 E_3\,$,
\begin{align}
&\;\,\hat{\alpha} = (1 + \cos\theta_{\rm min}) \Theta (p_3 - p_2) - 1\,, \qquad \hat{\beta} = (-1 + \cos\theta_{\rm max}) \Theta (p_3 - p_2) + 1\,, \nonumber \\
&\cos \theta_{\rm min,\, max} = \frac{1}{2 p_1 p_3} \left( - 2 \gamma - 2 p_2^2 - 2 m_\phi^2 \mp 2p_2 \sqrt{2\gamma + p_1^2 + p_2^2 + p_3^2 + 2 m_\phi^2}\, \right)\,,
\end{align}
where $\Theta$ is the Heaviside step function. The boundaries of the integral in $\cos\alpha$ are $x_{\mp} = \Big(- \hat{b} \pm \sqrt{\hat{b}^2 - 4 \hat{a} \hat{c}}\,\Big)\big/ (2\hat{a})$, with the definitions
\begin{align}
\hat{a} \,=&\; - 4 p_2^2 (p_1^2 + p_3^2 - 2 p_1 p_3 \cos\theta)\,, \nonumber\\
\hat{b} \,=&\; 8 p_2 (p_1 - p_3 \cos\theta) (p_1 p_3 \cos\theta + m_\phi^2 + \gamma)\,, \\
\hat{c} \,=&\; 4 \left[ p_2^2 p_3^2 ( 1 - \cos^2 \theta) - p_1^2 p_3^2 \cos^2 \theta - \gamma^2 - 2 p_1 p_3 \cos\theta\, \gamma - m_\phi^4 - 2 p_1 p_3 \cos\theta\, m_\phi^2 - 2 m_\phi^2 \gamma \right] \,. \nonumber
\end{align}
Notice that $\hat{a} \leq 0$. Finally, the squared matrix element is evaluated for $\cos\beta$ equal to
\begin{equation}
\cos\beta_+ = \frac{1}{p_2 p_3 \sin\alpha \sin\theta} \left(p_1 p_2 \cos\alpha - p_1 p_3 \cos\theta - p_2 p_3 \cos\alpha \cos\theta - m_\phi^2 - \gamma \right)\,.
\end{equation}

\subsection{Scattering mediated by the quartic coupling $\lambda_\phi$}\label{app:DRscatt_quartic}
First we consider the limit where $g_\phi$ is negligible and the matrix element squared in Eq.~\eqref{eq:general_Gamma} is simply $|\mathcal{M}|^2 = \lambda_\phi^2$. The integral over $\cos\alpha$ can then be performed analytically~\cite{Hannestad:1995rs},
\begin{equation}\label{eq:cosalpha_int}
\int_{x_-}^{x_+} \mathrm{d}\cos\alpha\, \frac{1}{\sqrt{\hat{a} \cos^2 \alpha + \hat{b} \cos\alpha + \hat{c}}} = \frac{\pi}{\sqrt{-\hat{a}}}\, \Theta (\hat{b}^2 -4 \hat{a}\hat{c})\,,
\end{equation}
where the $\Theta$ is in fact redundant, because in Eq.~\eqref{eq:general_Gamma} we already limited the $\cos\theta$ integration range accordingly. The resulting integration over $\cos\theta$ can also be performed analytically, leading us to
\begin{align}
&n_{\rm DR}^{(0)} \langle \Gamma_{\phi\phi}\rangle_{\lambda_\phi} = \frac{a}{16\pi^6}  \int_0^\infty \hspace{-1mm} \frac{\mathrm{d}p_1\hspace{0.2mm} p_1^2}{ 2E_1} \int_0^\infty \hspace{-1mm}\frac{\mathrm{d}p_2 \hspace{0.2mm} p_2^2}{2 E_2} \int_0^{p_{3}^{\rm max}} \hspace{-1mm} \frac{\mathrm{d} p_3 \hspace{0.2mm} p_3^2}{2 E_3}\, f_{\rm DR}^{(0)}(E_1) f_{\rm DR}^{(0)}(E_2) \big[1 + f_{\rm DR}^{(0)}(E_3)\big]  \nonumber\\ 
 \label{eq:quartic_only_Gamma}
&\times \big[1 + f_{\rm DR}^{(0)}(E_1+E_2 - E_3)\big] \left( \frac{-\,\pi \lambda_\phi^2}{4\hspace{0.2mm} p_1 p_2 p_3} \right) \left[ \sqrt{p_1^2 + p_3^2 - 2 p_1 p_3 \cos\theta}\, \right]^{\cos\theta\, =\, \hat{\beta}}_{\cos\theta\, =\, \hat{\alpha}}\;.
\end{align}
After changing integration variables to $x_i \equiv E_i/T_{\rm DR}$, it is straightforward to evaluate numerically the remaining $3$-dimensional integral.\footnote{We find that, after splitting the innermost integral as $\int_{0}^{p_2} \mathrm{d}p_3 + \int_{p_2}^{p_3^{\rm max}} \mathrm{d}p_3$, the two resulting terms give identical contributions. Hence we can evaluate only the $2\int_{0}^{p_2} \mathrm{d}p_3$ term, where the expression of the integrand is simpler, easing the numerical integration. The same applies to Eq.~\eqref{eq:cubic_only_Gamma} below.} For $m_\phi/ T_{\rm DR} \ll 1$, an accurate fit is given by
\begin{equation}
n_{\rm DR}^{(0)} \langle \Gamma_{\phi\phi}\rangle_{\lambda_\phi} \approx a\,6.5\, \frac{\lambda_\phi^2 T_{\rm DR}^4}{512 \pi^5} \bigg[ \log \frac{(4\pi T_{\rm DR})^2}{m_\phi^2} - 7.9\bigg]\,.
\end{equation}
This expression is valid at low temperatures, $T_{\rm DR} \ll T_{\rm c} = m_\phi/(\lambda_\phi/24)^{1/2}$. To extend its applicability to high temperatures~\cite{Kapusta:2006pm} we replace $m_\phi^2$ with the expression of the mass including thermal corrections, $m_\phi^2 + \lambda_\phi T_{\rm DR}^2/24$, obtaining
\begin{equation}\label{eq:Gammaphiphi_quartic_result}
\langle \Gamma_{\phi\phi}\rangle_{\lambda_\phi} \approx 6.5\hspace{0.2mm}\lambda_\phi^2\hspace{0.6mm} \frac{ T_{\rm DR, 0}}{512\hspace{0.2mm} \zeta(3) \pi^3}\, \Bigg[ \log \frac{(4\pi T_{\rm DR})^2}{m_\phi^2 +  \frac{\lambda_\phi T_{\rm DR}^2}{24}} - 7.9\Bigg] \,,
\end{equation}
where we have also used $n_{\rm DR}^{(0)} \simeq \zeta(3)T_{\rm DR}^3/\pi^2$. We see that at $T_{\rm DR} \gg T_{\rm c}$ the rate is temperature-independent (scaling as $\lambda_\phi^2 \log \big(16 \pi^2/\lambda_\phi\big)$, which is formally analogous to Debye screening for non-Abelian DR~\cite{Cyr-Racine:2015ihg,Rubira:2022xhb}). Therefore, once $\langle \Gamma_{\phi\phi}\rangle_{\lambda_\phi} = \mathcal{H}$ is reached the self-interaction remains coupled at all later times. For the values of the quartic coupling considered in this work, the DR recoupling happens early in radiation domination: for $\lambda_\phi = 10^{-5}\,(10^{-10})$  $\langle \Gamma_{\phi\phi}\rangle_{\lambda_\phi} = \mathcal{H}$ is reached at $z \approx 4 \times 10^{18}\,(7\times 10^8)$, assuming $\xi_0 = 0.5$. 

Equation~\eqref{eq:Gammaphiphi_quartic_result} improves upon the result of Refs.~\cite{Brinckmann:2022ajr,Chang:2025uvx}, which was obtained by neglecting Bose enhancement and adopting a Maxwell-Boltzmann form for the distribution, $f_{\rm DR}^{(0)}(E) = \pi^4 e^{-E/T_{\rm DR}}/90$. In that limit the integral on the right-hand side of Eq.~\eqref{eq:integral_definition_DRscatt} can be performed analytically~\cite{Gondolo:1990dk}, yielding $\langle \Gamma_{\phi\phi}\rangle^{(\mathrm{MB})}_{\lambda_\phi} \simeq \pi \lambda_\phi^2 \hspace{0.2mm} T_{\rm DR, 0}/23040$. For the values of $\lambda_\phi$ considered here, this is $30\,$-$\,60$ times smaller than our expression in Eq.~\eqref{eq:Gammaphiphi_quartic_result} at high redshift.

\subsection{Scattering mediated by the cubic coupling $g_\phi$}
As we showed in Appendix~\ref{app:DRscatt_quartic}, the self-scattering mediated by the quartic coupling~{\it alone} guarantees that we can describe DR as a fluid. Nevertheless, for completeness we discuss also the contribution of $g_\phi$, neglecting the interference with $\lambda_\phi$ for simplicity. The leading term in Eq.~\eqref{eq:general_Gamma} is then the sum of the squares of the $t$- and $u$-channel amplitudes,
\begin{equation}
|\mathcal{M}|^2 \simeq g_\phi^4 \Bigg[ \bigg(\frac{1}{t - m_\phi^2}\bigg)^2 + \bigg(\frac{1}{u - m_\phi^2}\bigg)^2 \Bigg] \quad \to\quad  2\hspace{0.2mm}g_\phi^4 \bigg(\frac{1}{t - m_\phi^2}\bigg)^2\,,
\end{equation}
where in second step we have exploited the fact that the two terms must contribute equally, due to Bose symmetry. Since $t - m_\phi^2 = m_\phi^2 - 2 (E_1 E_3 - p_1 p_3 \cos\theta)$, the integration over $\cos\alpha$ is performed immediately with the help of~\eqref{eq:cosalpha_int} and the integral over $\cos\theta$ can also be carried out analytically, leading us to
\begin{align}
&n_{\rm DR}^{(0)} \langle \Gamma_{\phi\phi}\rangle_{g_\phi} \simeq \frac{a}{16\pi^6}  \int_0^\infty \hspace{-1mm} \frac{\mathrm{d}p_1\hspace{0.2mm} p_1^2}{ 2E_1} \int_0^\infty \hspace{-1mm}\frac{\mathrm{d}p_2 \hspace{0.2mm} p_2^2}{2 E_2} \int_0^{p_{3}^{\rm max}} \hspace{-1mm} \frac{\mathrm{d} p_3 \hspace{0.2mm} p_3^2}{2 E_3}\, f_{\rm DR}^{(0)}(E_1) f_{\rm DR}^{(0)}(E_2) \big[1 + f_{\rm DR}^{(0)}(E_3)\big]  \nonumber\\ 
 \label{eq:cubic_only_Gamma}
&\times \big[1 + f_{\rm DR}^{(0)}(E_1+E_2 - E_3)\big]  \bigg( \frac{ -\, \pi g_\phi^4}{4\hspace{0.2mm} p_1 p_2 p_3 \big(2 E_1 E_3 - m_\phi^2 - p_1^2 - p_3^2 \big)} \bigg)   \\ 
&\times \left[  \frac{   \sqrt{p_1^2 + p_3^2 - 2  p_1 p_3 \cos\theta }}{ 2 E_1 E_3  - m_\phi^2 - 2 p_1 p_3 \cos\theta } + \frac{\arctan  \frac{\sqrt{p_1^2 + p_3^2 - 2 p_1 p_3 \cos\theta}}{\sqrt{2 E_1 E_3 - m_\phi^2 -p_1^2 - p_3^2}}\,}{\sqrt{2 E_1 E_3 - m_\phi^2 - p_1^2 - p_3^2}}\, \right]^{\cos\theta\, =\, \hat{\beta}}_{\cos\theta \,=\, \hat{\alpha}}\;. \nonumber
\end{align}
After changing variables to $x_i = E_i/T_{\rm DR}$, the triple integral is evaluated numerically. For $m_\phi/T_{\rm DR} \ll 1$ we obtain the fit
\begin{equation}\label{eq:Gammaphiphi_partial_g}
n_{\rm DR}^{(0)} \langle \Gamma_{\phi\phi}\rangle_{g_\phi}\approx a\, (0.9 \pm 0.3)\, \frac{g_\phi^4}{512 \pi^5}\, \bigg(\frac{T_{\rm DR}}{m_\phi}\bigg)^4\,,
\end{equation}
with a mild uncertainty on the overall coefficient, which was estimated by comparing different methods for the numerical integration. The expression~\eqref{eq:Gammaphiphi_partial_g} is valid for $T_{\rm DR}\ll T_{\rm c}$, whereas at $T_{\rm DR} \gg T_{\rm c}$ we must make the replacements $m_\phi^2 \to m^2_{\widehat{\phi},T}$ and $g_\phi \to g_{\widehat{\phi},T}\,$, where the high-temperature expressions of the mass and cubic coupling were given in Eq.~\eqref{eq:mass_cubic_highT}. In summary, we arrive at
\begin{equation} \label{eq:Gammaphiphi_cubic_result}
\langle \Gamma_{\phi\phi}\rangle_{g_\phi}\, \approx\, (0.9 \pm 0.3) \left( \frac{g_\phi}{m_\phi}\right)^4 \frac{T_{\rm DR,0}}{512 \zeta (3) \pi^3} \times  \begin{cases} 
      \,1 &\quad T_{\rm DR} \ll T_{\rm c}\,, \\
      \big(T_{\rm c}/T_{\rm DR})^{12} &\quad T_{\rm DR} \gg T_{\rm c}\,,
   \end{cases}
\end{equation}
where in the high-temperature limit we have taken $g_{\widehat{\phi},T} \simeq g_\phi T_{\rm c}^2/T_{\rm DR}^2$ as reasonable approximation, since $g_\phi^2 < 3\lambda_\phi m_\phi^2$ according to Eq.~\eqref{eq:gphi_max}. A comparison of Eqs.~\eqref{eq:Gammaphiphi_cubic_result} and~\eqref{eq:Gammaphiphi_quartic_result} shows that the contribution of $g_\phi$ to the DR self-scattering rate is at most comparable to the one of $\lambda_\phi$ at low temperatures, whereas at high temperatures $\langle \Gamma_{\phi\phi}\rangle_{g_\phi}$ is very steeply suppressed by thermal effects, with scaling $(T_{\rm c}/T_{\rm DR})^{12}$.

\section{Details of the Linear Perturbation Theory Computation}
\label{app:linear}
In this appendix we provide additional analytical formulae for the linear evolution of the density perturbations. We aim to solve Eq.~\eqref{eq:deltam}, repeated here for convenience,
\begin{align}\label{eq:deltam_app}
 \delta_m '' + \frac{3}{2y}\delta_m'-\frac{3}{2 y^2 }\delta_m-\frac{f_\chi\epsilon}{y^{1/2}}\left(\frac{3}{4}\delta_{\rm DR}'-\delta_\chi'\right)=0\,,
\end{align}
to first order in $\epsilon$. This implies it is sufficient to evaluate the source term to $\mathcal{O}(\epsilon)$. Making use of the analytical approximation for the DR density perturbation, Eq.~\eqref{eq:deltaDR_appr}, the solution to Eq.~\eqref{eq:deltam_app} is found to be 
\begin{align}
&\,\delta_m(k,y) =   \Bigg[ 1 - \frac{3\epsilon f_\chi}{50}  y^{1/2} Y(k,y)  \left(4 - 5 \log \frac{3Y(k,y)}{4+3 Y(k,y)}\right) + \frac{9\epsilon f_\chi}{160} \frac{Y(k,y)^2}{y^{3/2}} \bigg(4 y^2(1-Y(k,y)) \notag \\
   & +\frac{9}{2} y^2Y(k,y)^2- \frac{27}{4} y^2 Y(k,y)^3 - \frac{81}{16} y^2  Y(k,y)^4 \log \frac{3Y(k,y)}{4+3 Y(k,y)}
   \bigg) \Bigg]\delta_m^{\Lambda\mathrm{CDM} + \Delta N_{\rm eff}}(k,y)\,, \label{eq:deltam_full}
\end{align}
using the same definitions as in Sec.~\ref{sec:cosmo} and neglecting terms of $\mathcal{O}(\epsilon^2)$. The two initial conditions to the differential equation~\eqref{eq:deltam_app} were set such that at matter-radiation equality the decaying mode $\propto y^{-3/2}$ can be neglected, while the growing mode $\propto y$ equals its value in a $\Lambda\mathrm{CDM}+\Delta N_{\rm eff}$ cosmology. This is clearly only an approximation of the true dynamics, but one that is sufficient for our purposes. Other choices are possible, e.g.~nulling the $\mathcal{O}(\epsilon)$ correction to both density and velocity at equality, leading to slightly different numerical coefficients in the analytical solution above, but with no significant difference in the comparison to \texttt{CLASS}. The two asymptotic limits of Eq.~\eqref{eq:deltam_full} are 
\begin{align}
    \delta_m(k,y) \overset{\strut k\, \ll\, k_{s,\mathrm{DR}}}{\longrightarrow} \Bigg[1 - \frac{3 \epsilon f_\chi}{50}  y^{1/2} Y(k,y) \left(4 - 5 \log \frac{3Y(k,y)}{4}\right)\Bigg] \delta_m^{\Lambda\mathrm{CDM} + \Delta N_{\rm eff}}(k,y)
\end{align}
and 
\begin{align} \label{eq:deltam_largek}
 \delta_m(k,y)\overset{\strut k\, \gg\, k_{s,\mathrm{DR}}}{\longrightarrow}\Bigg[1-\frac{2\epsilon f_\chi}{3}  
y^{1/2} \left(1- \frac{6}{7\hspace{0.2mm}Y(k,y)}\right)\Bigg] \delta_m^{\Lambda\mathrm{CDM}+\Delta N_{\rm eff}}(k,y)   \,.
\end{align}
An interesting observation that can be drawn from Eq.~\eqref{eq:deltam_largek} is that the scale dependence of the signal at large wavenumber grows even faster than the overall suppression of the matter power spectrum, $y$ vs.~$y^{1/2}$. A formula that approximately combines the above scalings and is in good agreement with the \texttt{CLASS} output is
\begin{align}
\label{eq:deltam_resum}
    \delta_m(k,y)\, \simeq\,  \frac{1 + \tfrac{7}{6} Y(k,y)}{1 + \tfrac{7}{6}  Y(k,y) \big(1 + \tfrac{2}{3} f_\chi \epsilon  y^{1/2} \big)}\;\delta_m^{\Lambda \mathrm{CDM}+\Delta N_{\rm eff}}(k,y)\,,
\end{align}
showing how the shape suppression of the matter power spectrum depends on the Jeans scale of DR, while the high-$k$ limit is simply determined by $\epsilon$, as DR is completely negligible on small scales.

\begin{figure}
    \centering
    \includegraphics[width=0.5\linewidth]{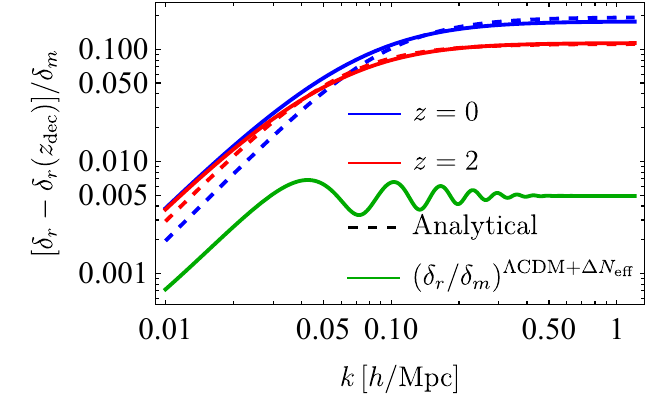}
    \caption{Relative density perturbation $\delta_r = \delta_b -\delta_\chi$ in a DM recoupling model, normalized to the total matter perturbation $\delta_m \equiv \delta_m^{\Lambda\mathrm{CDM}+\Delta N_{\rm eff}}$. We have set $\Delta N_{\rm eff} = 0.1$ and $a_{\mathrm{D}} = 100\;\mathrm{Mpc}^{-1}$. Solid~[dashed] lines correspond to \texttt{CLASS}~[our analytical solution in Eq.~\eqref{eq:deltar_sol_app}]. For ease of presentation an initial condition at CMB decoupling $z_{\rm dec}$, independent of the value of $a_{\rm D}$, has been subtracted. The shape of the relative density transfer function at $z = 0$ in a $\Lambda\mathrm{CDM}+\Delta N_{\rm eff}$ Universe is shown in green, for reference. For all practical purposes $\delta^{\Lambda \mathrm{CDM}+ \Delta N_{\rm eff} }_r(k,y) \simeq \delta_r^{\Lambda \mathrm{CDM} + \Delta N_{\rm eff}}(k,y_{\rm dec})$.}
    \label{fig:delta_r}
\end{figure}

Within our analytical approximations, we can also compute the relative density perturbation between baryons and DM, $\delta_r \equiv \delta_b -\delta_{\chi}$. In a $\Lambda$CDM Universe, relative density fluctuations do not grow after the redshift of baryon-photon decoupling $z_{\rm dec}$, and any residual relative velocity decays over time \cite{Schmidt:2016coo,Chen:2019cfu}. The shape of the transfer function of $\delta_r$ in a $\Lambda\mathrm{CDM}+ \Delta N_{\rm eff}$ model is shown by the green line in Fig.~\ref{fig:delta_r}, indicating that $\delta_r$ is less than a percent of the total matter fluctuation. In a model where DM recouples to DR, instead, the evolution of the two non-relativistic matter components is different, leading to growing relative perturbations. Under the same approximations employed above and in Sec.~\ref{sec:cosmo}, the relative density perturbation satisfies the equation
\begin{align}\label{eq:deltar_app}
 \delta_r '' + \frac{3}{2y}\delta_r' + \frac{\epsilon}{y^{1/2}}\left(\frac{3}{4}\delta_{\rm DR}'-\delta_\chi'\right)=0\,,
\end{align}
solved by
\begin{align}\label{eq:deltar_sol_app}
  \delta_r (k,y) =  \frac{27\hspace{0.2mm}\epsilon}{16} y^{1/2} Y(k,y) \Bigg[\frac{4}{27}+
    Y(k,y)^2  + \frac{1}{4} Y(k,y)^2 \big(2 + 3 Y(k,y)&\big)  \log \frac{3Y(k,y)}{4+3 Y(k,y)} \Bigg] \nonumber \\ &\times \delta_m^{\Lambda\mathrm{CDM} + \Delta N_{\rm eff}} (k,y)\,,
\end{align}
in terms of the total matter perturbation in a $\Lambda\mathrm{CDM}\,+\,\Delta N_{\rm eff}$ Universe and
neglecting $\mathcal{O}(\epsilon^2)$ corrections. The initial conditions were chosen such that both the decaying mode $\propto y^{-1/2}$ and the constant mode are negligible at matter-radiation equality. The agreement of the analytical formula with the \texttt{CLASS} solutions is quite remarkable, as displayed in Fig.~\ref{fig:delta_r}, which shows that relative density perturbations are more than twenty times larger in a recoupling model with $\Delta N_{\rm eff} = 0.1$ and $a_{\rm D} = 100\;\mathrm{Mpc}^{-1}$ than in a standard cosmological scenario. At small scales, Eq.~\eqref{eq:deltar_sol_app} takes the form
\begin{align}
       \delta_r(k,y) \overset{\strut k\, \gg\, k_{s,\mathrm{DR}}}{\longrightarrow}\frac{\epsilon}{3}\hspace{0.3mm} y^{1/2}  \left(1 - \frac{6}{5Y(k,y)}\right) \delta_m^{\Lambda\mathrm{CDM} + \Delta N_{\rm eff}} (k,y)\,,
\end{align}
indicating that, in this limit, relative perturbations do not depend on the fraction of interacting matter $f_\chi$. This is reminiscent of the behavior of relative perturbations in other scenarios with new physics in the dark sector \cite{Archidiacono:2022iuu,Bottaro:2023wkd}. 
Similar conclusions apply to the relative velocity divergence $\theta_r \equiv \theta_b - \theta_\chi$.

\section{More on the Phenomenology of the Microscopic Model}\label{app:Thermal_details}
In this appendix we collect additional results for the microscopic model defined in Eq.~\eqref{eq:Lag}. First, we summarize the derivation of the thermal effective potential employed in Sec.~\ref{sec:implications_model}. 
Second, we discuss the annihilation cross section relevant for the DM relic abundance. We then summarize the computation of the DM relic density from non-relativistic freeze out within the dark sector. Last, we discuss the ellipticity of galactic DM halos, which is usually reported as giving the strongest astrophysical constraint on long-range DM self-interactions mediated by light fields such as our $\phi$.

\subsection{Finite-temperature scalar potential}\label{sec:thermal_pot}
Using the formalism of finite-temperature field theory (see e.g.~Refs.~\cite{Quiros:1999jp,Laine:2016hma} for pedagogical reviews), the effective potential at 1-loop order and finite temperature $T$ can be written as~\cite{Dolan:1973qd,Weinberg:1974hy}
\begin{align}
    V_{\rm eff} (\phi, T) = V_{\rm tree} (\phi) + V^{T\,=\,0}_{1\text{-}\mathrm{loop}}(\phi) + V^{T\,>\,0}_{1\text{-}\mathrm{loop}} (\phi)\,,
\end{align}
where $V_{\rm tree}(\phi)$ is the tree-level scalar potential, $V^{T\,=\,0}_{1\text{-}\mathrm{loop}}(\phi)$ denotes the zero-temperature 1-loop correction and $V^{T\,>\,0}_{1\text{-}\mathrm{loop}}(\phi)$ the finite-temperature contribution (which vanishes in the limit $T\to0$).

We begin by discussing the tree-level potential $V_{\rm tree}$. We choose $g_{\phi}>0$ without loss of generality and impose $\lambda_{\phi} >0$ for stability. For $g_{\phi}^2 < 8 \lambda_{\phi} m_\phi^2/3$ only one minimum exists, at the origin in field space.\footnote{For $2 \lambda_{\phi} m_\phi^2 < g_\phi^2  < 8 \lambda_{\phi} m_\phi^2/3$ two distinct inflection points exist, located at $\phi < 0$.} In the narrow sliver $8 \lambda_{\phi} m_\phi^2/3 < g_\phi^2 < 3 \lambda_{\phi} m_\phi^2$ three stationary points exist, two of which are minima located at $\phi_0$ and $\phi_-$, where
\begin{equation}\label{eq:Vtree_minima}
    \phi_0 = 0 \,, \qquad \,\phi_{\mp} = -\,\frac{3}{2} \frac{ g_{\phi}}{ \lambda_{\phi}}\Bigg[ 1 \pm \bigg(1- \frac{8}{3}\frac{ \lambda_{\phi} m^2_{\phi} }{g^2_{\phi}}\bigg)^{1/2}\Bigg]\,,
\end{equation}
but $\phi_0$ is still the global minimum. %
%the minima are degenerate for $g_{\phi}^2 = 3 \lambda_{\phi} m_{\phi}^2$.
For $g_\phi^2 > 3 \lambda_\phi m_\phi^2$ the same stationary points~\eqref{eq:Vtree_minima} are present, but the global minimum is at $\phi_{-} < 0 $ (for $0 < m_\phi^2 < g_\phi^2/(3\lambda_\phi)$ the second minimum is at $\phi_0$, whereas for $m_\phi^2 < 0$ it is located at $\phi_+>0$). For simplicity, in this work we focus on the region defined by Eq.~\eqref{eq:gphi_max}, where the global minimum is at $\langle \phi \rangle = 0$ and the physical scalar excitation at zero temperature can be identified directly with $\phi$. In turn, this enforces $m_\phi^2 > 0$. In Sec.~\ref{sec:gphi2_large} below we discuss the complementary region $g_\phi^2 > 3 \lambda_\phi m_\phi^2\,$, showing that its predictions are not qualitatively different.

The zero-temperature correction to the potential, $V^{T\,=\,0}_{1\text{-}\mathrm{loop}}$, corresponds to the zero-momentum component of the effective action and is obtained by summing all 1-particle-irreducible 1-loop diagrams, with vanishing external momentum and arbitrary insertions of the background field $\phi$~\cite{Coleman:1973jx}. Explicitly,
\begin{equation}\label{eq:CW}
     V^{T\,=\,0}_{1\text{-}\mathrm{loop}}(\phi)= \sum_{i \,=\,\chi,\, \phi} (- 1)^{F_{i}} \frac{g_{[i]}}{2} \int \frac{\mathrm{d}^4 k_{\rm E}}{(2 \pi)^4}\log{\big[k_{\rm E}^2 + m_{i}^2(\phi) \big]}\,,
\end{equation}
where $k_{\rm E}^\mu$ denotes the Euclidean $4$-momentum of the $i^{\rm th}$-particle in the loop, $g_{[\chi]}= 4$ and $g_{[\phi]} = 1$ count the internal degrees of freedom, $F_\chi = 1~(F_\phi = 0)$ corresponds to fermionic~(bosonic) statistics, and $m_i(\phi)$ are the field-dependent masses in Eq.~\eqref{eq:field_dep_masses}. Equation~\eqref{eq:CW} is UV divergent and must be renormalized. We adopt cutoff regularization and impose on-shell renormalization conditions that preserve the tree-level values of all the parameters and of the vacuum expectation value $\langle\phi\rangle$, ensuring that the 1-loop minimum coincides with the tree-level one~\cite{Anderson:1991zb,Quiros:1999jp}.

The temperature-dependent correction $V^{T\,>\,0 }_{1\text{-}\mathrm{loop}} (\phi)$ is given, in the non-interacting gas approximation, by~\cite{Quiros:1999jp} 
\begin{align}
    {V}_{1\text{-}\mathrm{loop}}^{T\,>\,0} (\phi) =  \sum_{i \,=\,\chi, \phi} (- 1)^{F_{i}} \frac{g_{[i]}  T_{i}^4}{2 \pi^2}J_{B/F}\left(\frac{m_i^2(\phi)}{T_{i}^2}\right)\,,\label{eq:free_energy}
\end{align}
where we defined the thermal functions
\begin{equation}
      J_{B/F} (y^2) \equiv \int_0^{+\infty} \mathrm{d}x \, x^2 \log{\left(1\mp  e^{-\sqrt{x^2 + y^2}}\,\right)} = -\, y^2 \sum_{n\, =\, 1}^\infty \frac{(\pm 1)^n}{n^2}\, K_2 (ny) \,.
\end{equation}
We are interested in the regime where DM is highly non-relativistic, $T_\chi \ll m_\chi(\phi)$. Then $J_F$ admits a Boltzmann-suppressed expansion $\propto e^{-m_\chi(\phi)/T_{\chi}}$, as can be readily obtained from its expression in terms of Bessel functions, and can be safely neglected. Conversely, for $J_B$ we can employ a high-temperature expansion, whose leading terms are
\begin{align}\label{eq:unimproved_pot}
    V_{1\text{-}\mathrm{loop}}^{T\,>\,0}  (\phi) &\simeq -\,\frac{\pi^2}{90 }T_{\rm DR}^4 + \frac{T_{\rm DR}^2}{24} m^2_{\phi}(\phi) - \frac{T_{\rm DR}}{12 \pi} \big[m^2_{\phi}(\phi)\big]^{3/2}\,,
\end{align} 
where pieces that are not enhanced by $T_{\rm DR}$ were neglected. At temperatures $T_{\rm DR} \gg m_\phi(\phi)$, however, the naive weak-coupling expansion of the effective potential breaks down, due to the large hierarchy of scales. A signal of this breakdown is the appearance of the $\sim [m_\phi^2(\phi)]^{3/2}$ non-analytic term. In this regime, higher-loop diagrams become parametrically comparable to the 1-loop contribution~\cite{Laine:2016hma}. The dominant effect comes from the so-called daisy diagrams, which invalidate the perturbative expansion when $\lambda_{\phi}T_{\rm DR}^2/m_{\rm eff}^2 \sim 1$, where $m_{\rm eff}^2$ corresponds to the field-dependent scalar mass $m^2_{\phi}(\phi)$ in Eq.~\eqref{eq:field_dep_masses}. Hence, the perturbative expansion for the theory in Eq.~\eqref{eq:unimproved_pot} breaks down at the critical temperature $T_{\rm c } \sim m_{\phi}(\phi)/\lambda_{\phi}^{1/2}$.

This problem, which is due to the infrared sensitivity of bosonic thermal loops, can be cured by resumming the daisy diagrams~\cite{Parwani:1991gq, Arnold:1992rz,Curtin:2016urg}. To leading order in powers of the temperature, this amounts to replacing
\begin{equation}
m_\phi^2(\phi) \rightarrow m_\phi^2(\phi) + \frac{\lambda_\phi T_{\rm DR}^2}{24}
\label{eq:daisy_resum}
\end{equation}
in Eq.~\eqref{eq:unimproved_pot}. The complete renormalized potential, including both zero-temperature and daisy-resummed finite temperature corrections, is given in Eq.~\eqref{eq:V_thermal}, where some field-independent pieces were omitted. The expressions of the mass, cubic and quartic coupling of the physical excitation $\widehat{\phi}$ at $T_{\rm DR} \gg T_{\rm c}$ are given in and after Eq.~\eqref{eq:mass_cubic_highT}, whereas the higher-point interactions are strongly suppressed,
\begin{align}
\lambda_{\widehat{\phi}^{2q - 1}, T} \sim g_\phi\,\frac{ 3 \lambda_\phi m_\phi^2 - g_\phi^2}{\pi \lambda_{\phi}^{3/2} T_{\rm DR}^{2(q-1)}} \Bigg[ 1 + \mathcal{O} \bigg( \frac{T_{\rm c}^2}{T_{\rm DR}^2}\bigg) \Bigg]\,,\qquad
\lambda_{\widehat{\phi}^{2q}, T} \sim \frac{\lambda_\phi^{3/2}}{\pi  T_{\rm DR}^{2(q-2)}} \Bigg[ 1 + \mathcal{O} \bigg( \frac{T_{\rm c}^2}{T_{\rm DR}^{2}}\bigg) \Bigg]\,,
\end{align}
valid for $q\geq 3$. In addition, the phase transition from $\langle \phi\rangle = 0$ at low temperature to $\langle \phi \rangle \simeq - g_\phi/\lambda_\phi$ at high temperature causes a small relative shift of the DM mass by $(g_\phi/m_\phi)(y_\chi/\lambda_\phi)(m_\phi/m_\chi) \ll 1$, where the strong suppression by the tiny ratio $m_\phi/m_\chi$ dominates.

Inspection of beyond-daisy diagrams that are not included in the resummation shows~\cite{Quiros:1999jp} that, for the improved theory in Eq.~\eqref{eq:V_thermal}, validity of the perturbative expansion requires $\lambda_\phi T_{\rm DR}/m_{\rm eff} \ll \pi$, where now $m_{\rm eff}^2$ corresponds to the right-hand side of Eq.~\eqref{eq:daisy_resum}. This condition is automatically satisfied at $T_{\rm DR}\gg T_{\rm c}$, since by assumption $\lambda_\phi$ is perturbative.

\subsubsection{The region $g_\phi^2 > 3\lambda_\phi m_\phi^2$}\label{sec:gphi2_large}
We pause briefly to justify our choice of omitting the parameter region $g_\phi^2 > 3\lambda_\phi m_\phi^2$ from the main discussion. In this region the global minimum of the tree-level potential is at $\langle \phi \rangle = \phi_- <0$, where $\phi_-$ was defined in Eq.~\eqref{eq:Vtree_minima}. Hence, we expand as $\phi = \langle \phi \rangle + \widetilde{\phi}$ in terms of the physical excitation $\widetilde{\phi}$. The key features can be understood by considering two asymptotic limits:
\begin{itemize}
    \item $g_\phi^2\gg 3\lambda_\phi | m_\phi^2|$, which can be realized for either positive or negative $m_\phi^2$. The vacuum expectation value is found to be $\langle \phi \rangle \simeq -\, 3 g_\phi/\lambda_\phi$ and the physical mass and cubic coupling are
    \begin{equation}
        m_{\widetilde{\phi}}^2 \simeq \frac{3}{2} \frac{g_{\phi}^2}{\lambda_\phi} \,, \qquad g_{\widetilde{\phi}} \simeq -\, 2 g_\phi\,, \qquad  \frac{g_{\widetilde{\phi}}}{m_{\widetilde{\phi}}} \simeq -\, \frac{2\sqrt{2}}{\sqrt{3}}  \lambda_{\phi}^{1/2}\,.
    \end{equation}
    \item $g_\phi^2\ll -\, 3\lambda_\phi  m_\phi^2$, possible only for $m_\phi^2 < 0$. Here the vacuum is at $\langle \phi \rangle \simeq  - \sqrt{6} \big(-m_\phi^2\big)^{1/2}/\lambda_\phi^{1/2}$ and we obtain 
     \begin{equation}
       m^2_{\widetilde{\phi}} \simeq -\,2 m_{\phi}^2\,, \qquad g_{\widetilde{\phi}} \simeq - \,\sqrt{6} \lambda_\phi^{1/2} \big(\hspace{-0.5mm}- m_\phi^2\big)^{1/2}\,, \qquad \frac{g_{\widetilde{\phi}}}{  m_{\widetilde{\phi}}} \simeq -\, \sqrt{3}\hspace{0.3mm}  \lambda_{\phi}^{1/2}\,.
    \end{equation}
\end{itemize}
We see that, although the physical cubic coupling and mass are controlled by different parameters in the two cases, their ratio is always $g_{\widetilde{\phi}}/m_{\widetilde{\phi}} \sim \lambda_{\phi}^{1/2}$, up to $\mathcal{O}(1)$ numbers. This is the same parametric scaling that was realized for $g_\phi/m_\phi$ close to the upper limit of the region $g_\phi^2 < 3\lambda_\phi m_\phi^2$ studied in this paper (see in particular the solid lines in Fig.~\ref{fig:micro_param_space}, which assumed $g_\phi/m_\phi = \lambda_\phi^{1/2}$). 

The finite-temperature behavior is also qualitatively unchanged: the critical temperature is found to be
\begin{equation}
T_{\rm c}^2 = 24 \frac{m_{\widetilde{\phi}}^2}{\lambda_\phi}\,,
\end{equation}
where $m_{\widetilde{\phi}}$ is the physical mass of the scalar at zero temperature. Hence, $\lambda_\phi$ still needs to satisfy Eq.~\eqref{eq:lambdaphi_upperbound} in order for the critical temperature to be sufficiently high that the DM recoupling dynamics happens at $T_{\rm DR} \ll T_{\rm c}$. We conclude that all the features of this parameter region were, in fact, already captured by our discussion of the $g_\phi^2 < 3\lambda_\phi m_\phi^2$ regime; in particular, they are represented by the $g_\phi/m_\phi = \lambda_\phi^{1/2}$ benchmark, including fine-tuning considerations.

\subsection{Dark matter annihilation cross section}\label{app:DM_ann}
In the microscopic model of Eq.~\eqref{eq:Lag}, the relic abundance of $\chi$ and $\bar{\chi}$ is determined by thermal freeze out of $\chi\overline{\chi}\to\phi\phi$ annihilations in the early Universe. The tree-level amplitude for $\chi(p_1)\bar\chi(p_2)\to\phi(p_3)\phi(p_4)$ reads
\begin{equation}
\mathcal{M} (\chi\overline{\chi} \rightarrow \phi \phi) = -\, \bar{v}(p_2)\left[ y_{\chi}^2\left( \frac{i\slashed{p}_3 + 2m_\chi}{t - m_\chi^2} + \frac{i\slashed{p}_4 +2m_\chi}{u-m_\chi^2}\right) + \frac{y_{\chi} g_{\phi}}{s-m_\phi^2} \right] u (p_1)\,,
\end{equation}
where initial-state (final-state) momenta are taken ingoing (outgoing) and the Mandelstam variables are defined as in Eq.~\eqref{eq:Mandelstam}. The leading contribution to the annihilation cross section times the relative velocity between two DM particles in the center-of-mass (cm) frame, $v_{\rm cm}= \vert \vec{v}_{1} -\vec{v}_{2} \vert_{\rm cm}$, is~\cite{Gondolo:1990dk} 
\begin{equation}\label{eq:annichilation_crossec}
     (\sigma v_{\rm cm})_0 \simeq \frac{y_{\chi}^4  v_{\rm cm}^2}{384 \pi  m_\chi^2} \Bigg[\bigg(1 - \frac{g_{\phi}}{4m_\chi y_{\chi}} \bigg)^2 + 8 \bigg(1 - \frac{g_{\phi}}{8 m_\chi y_{\chi}} \bigg)^2\Bigg] \,,
\end{equation}
where we have expanded for small velocities and neglected $m_\phi$. The $p$-wave suppression observed in Eq.~\eqref{eq:annichilation_crossec} is a consequence of parity conservation~\cite{An:2016kie}. Since the ratio $g_{\phi}/(m_{\chi} y_{\chi}) \sim (g_{\phi}/m_{\phi})/y_{\chi} (m_{\phi}/m_{\chi}) \ll 1$ is strongly suppressed by the ratio between the DR and DM masses, we retain only the leading $\mathcal{O}\big(y_\chi^4\big)$ term in $(\sigma v_{\rm cm})_0$.

Beyond the above perturbative result, the exchange of the light scalar mediator $\phi$ effectively generates a Yukawa force between the DM particles, resulting in the Sommerfeld enhancement (SE) of the annihilation cross section~\cite{An:2016kie}. For $p$-wave annihilation in the relevant Coulomb regime $v_{\rm cm} \gg m_{\phi}/m_{\chi}$, the enhanced cross section reads $(\sigma v_{\rm cm})_{\rm SE} =\, (\sigma v_{\rm cm})_{\rm 0} \times S_1 (\alpha_{y} / v_{\rm cm})$, where~\cite{Cassel:2009wt} 
\begin{equation}
    S_1 (\alpha_{y}  / v_{\rm cm})=\frac{2\pi \alpha_{y} /v_{\rm cm}}{1 - \exp (-\,2\pi \alpha_{y} /v_{\rm cm})} \bigg[ 1 + \frac{(2 \alpha_{y} /v_{\rm cm})^2}{4}\bigg]\,, \label{eq:SE_enhancedcrosssection}
\end{equation}
and $\alpha_{y}  \equiv y_{\chi}^2/(4 \pi) $. For interaction strengths $\pi \alpha_{y}\gtrsim v_{\rm cm}$, the SE factor has a significant impact on the cross section, causing $(\sigma v_{\rm cm})_{\rm SE}$ to grow as $\sim 1/v_{\rm cm }$ at small velocity.
Assuming a Maxwell-Boltzmann velocity distribution for the DM particles, $f_{\chi}(\vec{v}_i\hspace{0.3mm})= e^{ - v_i^2/v_0^2} /(\pi^{3/2}v_0^3)$, the thermally-averaged annihilation cross section is found to be~\cite{Feng:2010zp}
\begin{equation}
    \langle \sigma v_{\rm cm}\rangle = \frac{\sqrt{2}}{\pi^{1/2} v_0^3} \int^{+\infty}_{0} \mathrm{d} v_{\rm cm} v_{\rm cm}^2 e^{- \frac{v_{\rm cm}^2}{2 v_0^2}}(\sigma v_{\rm cm })_{\rm SE}\,,    \label{eq:averaged_thermal_relic}
\end{equation}
where $v_0\equiv \sqrt{2 \, T_{\rm DR}/m_\chi}$ is the most probable velocity of the DM particles (DM and DR remain in kinetic equilibrium throughout the freeze-out process, hence $T_{\rm DR} = T_{\chi}$). Equation~\eqref{eq:averaged_thermal_relic} allows us to evaluate $\langle \sigma v_{\rm cm}\rangle$ for given values of $m_\chi, y_\chi$ and $T_{\rm DR}$. If $\chi$ and $\overline{\chi}$ acquire the observed DM abundance via freeze out, then $T_{\rm DR, fo} \sim m_\chi/ 25$, corresponding to $v_0 \sim 0.3$. Therefore, Sommerfeld enhancement significantly modifies the relic abundance only for $y_{\chi}\gtrsim 1$. This matches the point where the thermal relic abundance curve in Fig.~\ref{fig:micro_param_space} (shown in red) deviates from a straight line.

\subsection{Secluded freeze out}\label{app:DM_relic_density}
Here we express the present-day relic abundance of the $\chi$ species in terms of the thermally averaged annihilation cross section computed in Eq.~\eqref{eq:averaged_thermal_relic}. To do so, we generalize the discussion to the secluded freeze out of an arbitrary $2\to 2$ annihilation process $\chi\overline{\chi}\to XX$, assuming no asymmetry is present between $\chi$ and $\bar{\chi}$ particles. We assume that $\chi$ and $X$ remain in kinetic equilibrium well after the former freezes out, as realized for a broad set of possible interactions mediating the annihilation and scattering processes, including the model studied in this paper.

Denoting with $Y_{\chi} \equiv n^{(0)}_\chi/s$ the ratio between the background number density of $\chi$ and the total entropy density, the evolution of $Y_{\chi}$ during radiation domination is governed by the Boltzmann equation~\cite{Gondolo:1990dk}
\begin{equation}
   \frac{d Y_{\chi}}{dx} = -\,\sqrt{\frac{8 \pi^2}{45}} \frac{g_{* s}}{\sqrt{g_*}} m_\chi M_{\rm Pl}\,\frac{\langle \sigma v_{\rm M\o l}\rangle}{x^2}  \bigg[Y_{\chi}(x)^2 - Y_{\chi, \mathrm{eq}}(x)^{2}\bigg]\,,\label{eq:Boltzmann_appendix}
\end{equation}
where $x\equiv m_\chi/T_\gamma$, while $g_{\ast s}$~($g_*$) denotes the number of effective relativistic degrees of freedom for the entropy~(energy) density, and we neglected a small correction due to the variation of $g_{\ast s}$ with temperature. We adopt a general parametrization of the thermally-averaged annihilation cross section,
\begin{equation}
    \langle \sigma v_{\rm M\o l}\rangle= \sigma_0 \, \xi^{n}x^{-n}\,,
\end{equation}
where $\xi =T_{\rm DR}/T_{\gamma}$ is the ratio between the dark sector and photon temperatures and $n=0$ for $s$-wave annihilation, $n=1$ for $p$-wave, and so on. In the non-relativistic limit, the equilibrium number density $n^{(0)}_{\chi, \rm eq}$ follows a Maxwell-Boltzmann distribution, hence
\begin{equation}
Y_{\chi, \mathrm{eq}}(x) = K\xi^{3/2} x^{3/2} e^{-x/\xi},
\qquad
K = \frac{45\sqrt{\pi}}{4\sqrt{2}\pi^4}\frac{\eta_\chi}{g_{\ast s}}\,.
\end{equation}
The freeze-out temperature $x_{\rm fo} = m_{\chi}/T_{\gamma,\rm fo}$ corresponds to the time when $Y_\chi$ departs from $Y_{\chi,\rm eq}$, which can be estimated from the implicit condition~\cite{Scherrer:1985zt}
\begin{equation}
x_{\rm fo}^{\,n+\frac{1}{2}}
e^{x_{\rm fo}/\xi_{\rm fo}}
=
(2+\delta)\delta\hspace{0.3mm} K \lambda\hspace{0.5mm} \xi_{\rm fo}^{\,n+\frac{5}{2}}\,, \qquad \lambda = \sqrt{\frac{8 \pi^2}{45}}  \frac{g_{* s}}{\sqrt{g_*}} m_\chi M_{\rm Pl}\hspace{0.3mm} \sigma_0 \,,
\label{eq:xfo_master_appendix}
\end{equation}
where $\delta = 1 \,(2)$ for $s$-wave ($p$-wave) annihilation (see e.g.~Ref.~\cite{Chen:2013bi}) is a parameter included to match the exact numerical solution of Eq.~\eqref{eq:Boltzmann_appendix}; note that $x_{\rm fo}$ is only logarithmically sensitive to the choice of $\delta$. In evaluating $x_{\rm fo}$, we neglect any possible SE in the thermally-averaged cross section. Solving Eq.~\eqref{eq:xfo_master_appendix} iteratively, we find an analytic solution for the freeze-out temperature,
\begin{equation}
\frac{x_{\rm fo}}{\xi_{\rm fo}}
\,\simeq\, \log\!\left((2+\delta)\delta\hspace{0.3mm} K \lambda\,\xi_{\rm fo}^{\,n+\frac{5}{2}}\right) - \left(n+1/2\right) \log\!\left[ \xi_{\rm fo} \log\!\left((2+\delta)\delta\hspace{0.3mm} K \lambda\,\xi_{\rm fo}^{\,n+\frac{5}{2}}\right) \right]\,.
\label{eq:xfo_solution_appendix}
\end{equation}
Equations~\eqref{eq:xfo_master_appendix} and~\eqref{eq:xfo_solution_appendix} exhibit a slightly different scaling with $\xi$ compared to previous literature~\cite{Agrawal:2016quu, Feng:2008mu}. In Eq.~\eqref{eq:xfo_solution_appendix}, $\xi_{\rm fo}$ is related to the present value $\xi_0$ by entropy conservation,
\begin{equation}\label{eq:xi_fo}
  \xi_{\rm fo} =  \bigg(\frac{g_{\ast s\, \mathrm{vis}, \rm fo}}{g_{\ast s\,\mathrm{vis}, 0}}\bigg)^{1/3}\xi_{\rm 0}\,,
\end{equation}
where the contribution of $\chi$ to the dark sector entropy is Boltzmann suppressed and therefore negligible at freeze out.\footnote{In practice, we solve Eq.~\eqref{eq:xi_fo} for $\xi_{\rm fo}$ by making the ansatz $T_{\gamma, \mathrm{fo}} \to m_\chi /(25\hspace{0.3mm} \xi_{\rm fo})$ in $g_{\ast s\, \mathrm{vis}, \rm fo}$. We then plug the value of $\xi_{\rm fo}$ into Eq.~\eqref{eq:xfo_solution_appendix} and compute $x_{\rm fo}$, assuming again $T_{\gamma, \mathrm{fo}} \to m_\chi /(25\hspace{0.3mm} \xi_{\rm fo})$ in the evaluation of $g_\ast$.}

The relic density can be computed by solving Eq.~\eqref{eq:Boltzmann_appendix} in the late-time regime $x\gg x_{\rm fo}$, where $Y_{\chi, \rm eq}(x)$ is negligible. One obtains~\cite{Gondolo:1990dk}
\begin{align}
   \, \Omega_{\rm \chi + \bar{\chi}}  = 2\hspace{0.3mm}\frac{ s_0 Y_{\chi}(\infty)m_\chi}{\rho_{\rm cr}}= 2\hspace{0.3mm}\frac{ \pi  (n +1)}{9 \sqrt{10} }\frac{g_{* s, 0}}{M_{\rm Pl}^3}\frac{T_{\gamma, 0}^3}{H_0^2}\frac{\sqrt{g_{*, \mathrm{fo}}}}{g_{* s,\mathrm{fo}}}\frac{x_{\rm fo}}{\langle{\sigma v_{\rm M\o l}\rangle}_{\rm{fo}}}\,,
   \label{eq:DM_relic}
\end{align}
where $s_0$ denotes the entropy density today. 

In the model studied in this paper, DM annihilation is a $p$-wave process, hence we employ the above formulas with $n = 1$. For $\langle \sigma v_{\rm M\o l}\rangle_{\rm fo}$ we use the expression in Eq.~\eqref{eq:averaged_thermal_relic} evaluated at freeze out. The thermal relic abundance curve shown in red color in Fig.~\ref{fig:micro_param_space} is obtained by requiring $\Omega_{\chi + \bar{\chi}}\hspace{0.3mm}h^2 = 0.120$. Notice that in this paper $\Omega_\chi$ denotes the energy density of {\it all} of DM.

\subsection{Constraint from ellipticity of dark matter halos}
A small-scale constraint on the parameter space we consider in Sec.~\ref{sec:implications_model} comes from the observed ellipticity of galaxy halos~\cite{Buote:2002wd}, since DM self-interactions mediated by a light boson (like our $\phi$) can reduce the degree of anisotropy of the halo DM velocity distribution through many soft scatterings~\cite{Feng:2009mn,Agrawal:2016quu,McDaniel:2021kpq}. In the non-relativistic limit, the differential cross section for the scattering of two DM particles mediated by $\phi$ exchange is, in the center-of-mass frame,
\begin{equation}
\frac{d\sigma_{\chi \bar{\chi}\, \rightarrow\, \chi \bar{\chi}}}{d\Omega}  = \frac{4\alpha_y^2}{m_\chi^2 v_{\rm cm}^4 \Big( 1 - \cos\theta + \frac{2m_\phi^2}{m_\chi^2 v_{\rm cm}^2}\Big)^2}\,, 
\end{equation}
where $v_{\rm cm}= \vert \vec{v}_{1} -\vec{v}_{2} \vert_{\rm cm}$. This expression applies both to same-charge and opposite-charge scatterings. The constraint is estimated~\cite{Feng:2009mn} by requiring that the characteristic timescale $\tau_{\rm iso}$ for an average DM particle to change its kinetic energy by $\mathcal{O}(1)$, which is interpreted as the relaxation time needed to isotropize the DM velocity distribution, be longer than the age of the Universe. If the mediator mass is negligible, following Ref.~\cite{Agrawal:2016quu} we cut off the infrared ``Coulomb logarithm'' present in $\tau_{\rm iso}$ at the inter-particle spacing $\lambda_{\rm P} \equiv (m_\chi/\rho_\chi)^{1/3}$. Conversely, if the mediator is heavy the cutoff is provided by $m_\phi$. In summary, our calculation of the relaxation time gives
\begin{equation}
\tau_{\rm iso} \simeq \begin{cases}
\frac{3}{16\sqrt{\pi}} \frac{m_\chi^3 v_0^3}{\alpha_y^2 \rho_\chi \log \big( \frac{\lambda_{\rm P}^2 m_\chi^2 v_0^4}{\alpha_y^2}\big)} & \quad m_\phi < \frac{\alpha_y}{\lambda_{\rm P} v_0}\,, \\
\frac{3}{16\sqrt{\pi}} \frac{m_\chi^3 v_0^3}{\alpha_y^2 \rho_\chi \log \big( \frac{m_\chi^2 v_0^2}{m_\phi^2}\big)} & \quad m_\phi > \frac{\alpha_y}{\lambda_{\rm P} v_0}\,, \label{eq:tau_ISO}
\end{cases} 
\end{equation}
where $v_0$ is related to the velocity dispersion of DM particles by $\langle v^2 \rangle = 3v_0^2/2$. We estimate the bound by requiring $\tau_{\rm iso}> 10\;\mathrm{Gy}$, where $\rho_\chi$ and $v_0$ are set to typical values for the NGC720 galaxy studied by Ref.~\cite{Buote:2002wd}: we choose $\rho_\chi = 0.7\;\mathrm{GeV}/\mathrm{cm}^3$ and $\langle v^2 \rangle^{1/2} = 250\;\mathrm{km}/\mathrm{s}$, corresponding to a radius $r = 10\;\mathrm{kpc}$~\cite{Feng:2009mn}. Our estimate of the ellipticity bound is shown by the dashed  green line in Fig.~\ref{fig:micro_param_space}. The transition between the two regimes in Eq.~\eqref{eq:tau_ISO} happens for $m_\chi \approx 300\;\mathrm{GeV}$ (for $m_\chi$ larger~(smaller) than this, the first~(second) expression applies), but only amounts to a small change in slope. More recent work extended the ellipticity analysis to a sample of $11$ galaxies~\cite{McDaniel:2021kpq}, broadly confirming previous results. A thorough discussion of caveats to the ellipticity constraint can be found in Ref.~\cite{Agrawal:2016quu}.

\bibliographystyle{JHEP.bst}
\bibliography{ref.bib}

\end{document}